\documentclass[twocolumn,notitlepage,superscriptaddress,nofootinbib,tightenlines]{revtex4-1}
\usepackage{amsmath,verbatim,latexsym,amssymb,graphicx,indentfirst,mathrsfs,mathtools,amsthm,bbm,bm,hyperref,url,cancel,subcaption,yfonts}
\usepackage[font=small,labelfont=bf,format=plain,justification=raggedright,singlelinecheck=false]{caption}
\usepackage{verbatim,indentfirst}
\usepackage[title,titletoc]{appendix}
\usepackage[shortlabels]{enumitem}

\usepackage{color}
\usepackage[dvipsnames]{xcolor}
\usepackage[normalem]{ulem}

\usepackage{soul}

\newcommand{\Dim}{ d }  
\newcommand{\Sites}{n}  
\newcommand{\Sys}{{ \mathcal{S} }}  
\newcommand{\Control}{{ \mathcal{C} }}  
\newcommand{\Controll}{{ \mathcal{C}' }}  

\newcommand{\PQIM}{{\rm P}}
\newcommand{\FInt}[1]{{ F_{#1}^{\rm int} }}  
\newcommand{\VInt}{{ V_{\rm int} }}  
\newcommand{\FWeak}[1]{{ F_{#1}^{\rm wk} }}  
\newcommand{\SumKDWeak}[1]{{ \SumKD{#1}^{\rm wk} }}  
\newcommand*{\SumKD}[1]{\tilde{ \mathscr{A} }_{#1}}  
\newcommand{\W}{W}
\newcommand*{\ProjW}[1]{\Pi^{ \W }_{#1}}  
\newcommand*{\ProjV}[1]{\Pi^{ V }_{#1}}  

\newcommand*{\Depol}[1]{{\mathcal{E}_{#1}^{\rm depol}}}  
\newcommand*{\Depoll}[1]{{ \tilde{ \mathcal{E} }_{#1}^{\rm depol} }}  
\newcommand*{\FDepol}[3]{ {F_{{#1},{#2},{#3}}^{\rm depol}} }
\newcommand{\ideal}{{\rm ideal}}

\newcommand{\DegenV}{ \lambda }  

\newcommand{\reg}{ {\rm reg} } 

\newcommand{\Coupling}{{\rm c}}
\newcommand{\Left}{{\rm L}}
\newcommand{\Right}{{\rm R}}

\newcommand*{\ket}[1]{\lvert #1 \rangle}

\newcommand*{\ketbra}[2]{\lvert #1 \rangle\!\langle #2 \rvert}
\newcommand*{\expval}[1]{\left\langle  #1  \right\rangle}

\newcommand{\Tr}{{\rm Tr}}   
\def\id{\mathbbm{1}}   
\newcommand{\kB}{k_\mathrm{B}}  
\newcommand{\Hil}{\mathcal{H}}  

\newcommand{\LParen}{ \bm{(} }
\newcommand{\RParen}{ \bm{)} }

\renewcommand\th{ {\rm th} }

\usepackage{graphicx}
\usepackage{epstopdf}
\newcommand{\caphead}[1]{{\bf #1}}

\makeatletter
\newcommand\footnoteref[1]{\protected@xdef\@thefnmark{\ref{#1}}\@footnotemark}
\makeatother

\usepackage{tikz}
\newcommand{\Plus}{\mathord{\begin{tikzpicture}[baseline=0ex, line width=1, scale=0.13]
\draw (1,0) -- (1,2);
\draw (0,1) -- (2,1);
\end{tikzpicture}}}

\begin{document}
\title{Resilience of scrambling measurements}
\author{Brian~Swingle}
\affiliation{Condensed Matter Theory Center, Maryland Center for Fundamental Physics, Joint Center for Quantum Information and Computer Science, and Department of Physics, University of Maryland, College Park, MD 20742, USA}
\affiliation{Kavli Institute for Theoretical Physics, University of California, Santa Barbara, CA 93106, USA}
\author{Nicole~Yunger~Halpern\footnote{E-mail: nicoleyh@caltech.edu}}
\affiliation{Institute for Quantum Information and Matter, Caltech, Pasadena, CA 91125, USA}
\affiliation{Kavli Institute for Theoretical Physics, University of California, Santa Barbara, CA 93106, USA}

\begin{abstract}
Most experimental protocols for measuring scrambling require time evolution with a Hamiltonian and with the Hamiltonian's negative counterpart (backwards time evolution). Engineering controllable quantum many-body systems for which such forward and backward evolution is possible is a significant experimental challenge. Furthermore, if the system of interest is quantum-chaotic, one might worry that any small errors in the time reversal will be rapidly amplified, obscuring the physics of scrambling. This paper undermines this expectation: We exhibit a \emph{renormalization protocol} that extracts nearly ideal out-of-time-ordered-correlator measurements from imperfect experimental measurements. We analytically and numerically demonstrate the protocol's effectiveness, up to the scrambling time, in a  variety of models and for sizable imperfections. The scheme extends to errors from decoherence by an environment.
\end{abstract}

{\let\newpage\relax\maketitle}

%
%
\section{Introduction}
\label{section:Introduction}

Quantum information \emph{scrambles} when it spreads over all the degrees of freedom of a quantum many-body system, becoming inaccessible to few-body probes~\cite{Hayden_07_Black,Sekino_08_Fast,Brown_13_Scrambling}. In a recent spate of theoretical activity, scrambling has been related to early-time signatures of quantum chaos~\cite{LarkinO_69,Shenker_Stanford_14_BHs_and_butterfly,Kitaev_15_Simple,Maldacena_15_Bound}, to the scattering of high-energy quanta near a black-hole horizon~\cite{Dray_85_Shock,Shenker_Stanford_15_Stringy},
to bounds on the propagation of quantum information~\cite{Roberts_16_Lieb},
to quasiprobabilities (nonclassical generalizations of probabilities)~\cite{YungerHalpern_17_Jarzynski,NYH_17_Quasi},
to thermodynamic fluctuation relations~\cite{YungerHalpern_17_Jarzynski,Campisi_16_Thermodynamics,Tsuji_18_Out},
to Schwinger-Keldysh path integrals~\cite{Aleiner_16_Microscopic,Haehl_16_Schwinger_I,Haehl_16_Schwinger_II,Hael_17_Classification},
to quantum channels~\cite{HosurYoshida_16_Chaos},
to unitary $k$-designs~\cite{Roberts_16_Chaos,Cotler_17_Chaos,HunterJones_17_Chaos},
and to much else. On the experimental side, many proposals for observing scrambling now exist~\cite{Swingle_16_Measuring,Yao_16_Interferometric,Zhu_16_Measurement,Danshita_16_Creating,YungerHalpern_17_Jarzynski,NYH_17_Quasi,Bohrdt_16_Scrambling,Campisi_16_Thermodynamics,Tsuji_17_Exact}, and at least four early experiments have been performed~\cite{Li_16_Measuring,Garttner_16_Measuring,Wei_16_NuclearSpinOTOC,Meier_17_Exploring}.

Central to these developments is a physical quantity called
the \emph{out-of-time-ordered correlator} (OTOC).
Consider a quantum many-body system
governed by a Hamiltonian $H$
that generates the time-evolution unitary $U$.
Let $\rho$ denote a state of the system,
e.g., a thermal state $e^{ - \beta H } / Z$,
for some inverse temperature $\beta$ and
a partition function $Z$.
Let $W$ and $V$ denote Hermitian or unitary operators
defined on the system's Hilbert space.
$W$ evolves as $W_t := U^\dag W U$ in the Heisenberg picture.
The OTOC is defined as
\begin{align}
   F_t := \langle W^\dag_t V^\dag W_t V \rangle
   \equiv  \Tr (  W^\dag_t V^\dag W_t V  \rho ) \, .
\end{align}

The operators' ordering lends the OTOC its name.
We can grasp one significance of $F_t$ by assuming that
$\rho  =  \ketbra{ \psi }{ \psi }$ is pure,
$V$ is unitary, and $W$ is Hermitian.
Consider two protocols that differ just via an order of operations:
(i) Prepare $\ket{ \psi }$,
perturb the system with $V$, evolve the system forward in time with $U$,
measure $W$, and evolve the system backward with $U^\dag$.
This protocol prepares $W_t V \ket{ \psi }  =:  \ket{ \psi' }$.
(ii) Prepare $\ket{ \psi }$,
evolve the system forward, measure $W$, evolve the system backward,
and measure $V$.
This protocol prepares $V W_t \ket{ \psi }  =: \ket{ \psi'' }$.
The discrepancy between the protocols
imprints on the overlap $| \langle \psi'' | \psi' \rangle |   =  | F_t |$.

As this forward-and-backward explanation suggests, OTOCs resemble
the well-known Loschmidt echo in spirit (see \cite{Prosen_03_Echo_Review,Goussev_12_Echo_Review} for a review).
Like observations of the echo, most OTOC-measurement proposals require the experimenter to effectively reverse the flow of time.
Unfortunately, effective time reversal is typically experimentally challenging.
No general method for circumventing this difficulty is known.
OTOC-measurement protocols that do not require time reversal suffer from other limitations that likely preclude the study of large systems. 
Nevertheless, progress in the control of atoms, molecules, ions, and photons
has brought experimental measurements of OTOCs and scrambling seemingly within reach~\cite{Li_16_Measuring,Garttner_16_Measuring,Wei_16_NuclearSpinOTOC,Meier_17_Exploring}.

One may wonder if the difficulty of
precisely reversing time's flow
is more than technical.
Perhaps, for sufficiently large, complex, chaotic quantum many-body systems, small imperfections in the time-reversal procedure will always be amplified and obscure the physics of interest.
We believe that a fault-tolerant quantum computer could implement the time reversal with satisfactory accuracy. But do we need such a resource?

We argue that these concerns, while reasonable, are not borne out in practice.
We show how a simple renormalization procedure
can be used to extract OTOCs' early-time dynamics.
The renormalization requires only experimentally measurable quantities.
The dynamics of chaotic quantum many-body systems
can be recovered.

We offer theoretical arguments, and numerical and analytical evidence, for the following claim: The ideal OTOC's essential physics can, up to the scrambling time,
be extracted from imperfect measurements in which
the forward and backward time evolutions
differ by $10\%$ or more from their ideal forms:
Each implemented Hamiltonian differs from the ideal Hamiltonian $H$
by terms that carry an overall scale factor $\varepsilon \leq 0.1$.
This \textit{resilience} is quite universal:
The system can exhibit strong chaos or integrability.
The interactions can be local or nonlocal.
Our result holds even when imperfections vary from experimental run to experimental run.

Detailed numerical studies of a one-dimensional quantum Ising chain
support our general derivations.
So does an analytical calculation with
a strongly chaotic model dual to a black hole.
The renormalization scheme works here if
the time $t$ for which the system evolves forward differs from
the time $t''$ for which the system evolves backward.
Though Hamiltonian errors motivate much of this paper,
also decoherence by the environment threatens OTOC measurements.
The renormalization scheme helps combat decoherence,
as we show with numerical simulations and tailored analytical calculations.

Our physical picture of this resilience phenomenon
is that the imperfect OTOC contains two pieces of physics.
One piece consists of the growth of operators, and the spreading of information, characteristic of scrambling.
One piece consists of the decay of fidelity
due to mismatched forward and backward time evolutions
(similar to the traditional Loschmidt echo).
We claim that these two pieces of physics can be effectively separated, and that the second piece can be cleaned off from the first,
until the scrambling time, through the use of only experimentally measurable data.

We focus on two scrambling protocols, the interferometric protocol~\cite{Swingle_16_Measuring} and the weak-measurement protocol~\cite{YungerHalpern_17_Jarzynski,NYH_17_Quasi}.
But we expect our results to extend to other OTOC measurement schemes.
The paper is structured as follows:
Section~\ref{section:Interferometer} concerns the interferometric scheme.
Section~\ref{section:Weak} concerns the weak measurement scheme.
Section~\ref{section:Decoher} concerns
environmental decoherence (for both schemes).
Section~\ref{section:Holographic} shows our scheme's efficacy in
a strongly chaotic holographic model
plagued by unequal-time evolutions,
via analytical calculation.
Section~\ref{section:Conclusions} concludes with
future directions and open questions.

\section{Example \#1: Interferometer}
\label{section:Interferometer}

The interferometric scheme for measuring the OTOC
was introduced in~\cite{Swingle_16_Measuring}.
The set-up and protocol are reviewed in Sec.~\ref{section:Interf_Setup}.
The protocol can suffer from Hamiltonian errors
detailed in Sec.~\ref{section:Interf_ErrHam}.
The renormalization scheme mitigates those errors.
We motivate the renormalization mathematically
in Sec.~\ref{section:Interf_PseudoDerivn}.
Section~\ref{section:Interf_Numerics} supports the scheme
with numerical simulations of
the power-law quantum Ising model.

\subsection{Set-up and protocol for the interferometer}
\label{section:Interf_Setup}

Let $\Sys$ denote the system of interest,
associated with a Hilbert space $\Hil$.
We illustrate with a chain of $\Sites$ qubits
(spin-$\frac{1}{2}$ degrees of freedom).
Let $\sigma^\alpha_j$ denote
the $\alpha = x, y, z$ component of
the $j^\th$ site's spin.
The $+1$ and $-1$ eigenstates of $\sigma^z$
are denoted by $\ket{ 0 }$ and $\ket{1}$.

A Hamiltonian $H$ determines the system's natural dynamics.
$H$ generates the time-evolution operator
$U  :=  e^{ - i H t }$.

Let $W$ and $V$ denote local unitaries.
Unitaries that nontrivially transform only faraway subsystems
reflect scrambling.
For example, $W$ can manifest as
the first qubit's Pauli-$z$ operator:
$W  =  \sigma^z_1  \otimes  \id^{ \otimes  ( \Sites - 1 )}$.
$V$ can manifest as
the final qubit's Pauli-$x$ operator:
$V  =  \id^{ \otimes  ( \Sites - 1 )}  \otimes  \sigma^x_\Sites$.
In the Heisenberg Picture, $W$ evolves as
$W_t  :=  U^\dag  W  U$.

\begin{figure}[hbt]
\centering
\includegraphics[width=.45\textwidth, clip=true]{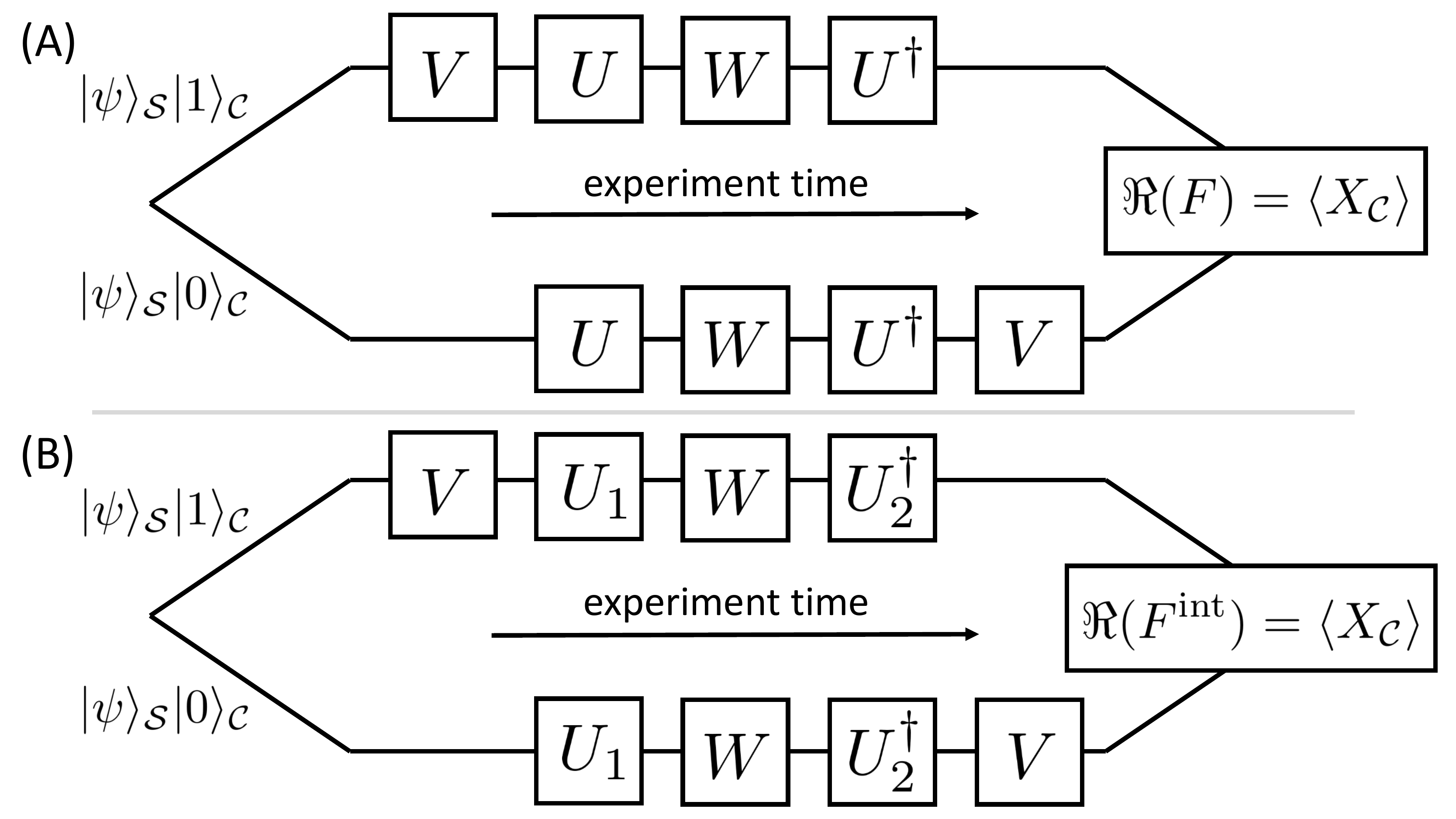}
\caption{\caphead{Interferometric protocol for measuring
the out-of-time-ordered correlator (OTOC):}
Panel (A) shows the ideal interferometer for measuring the OTOC in which
the forward ($U$) and backward ($U^\dagger$) evolutions are ideal:
$U=e^{-i H t}$, and $U^\dagger = e^{i H t}$.
Panel (B) shows the perturbed interferometer.
The forward evolution is $U_1 = e^{-i H_1 t}$, and
the backward evolution is $U_2^\dagger  = e^{i H_2 t}$.
A control qubit $\Control$ is initially prepared in the state
$\ket{ + }_\Control
=  \frac{1}{ \sqrt{2} }  ( \ket{ 0 }_\Control   +  \ket{ 1 }_\Control  )$.
The $\ket{0}_\Control$ defines one interferometer branch,
and the $\ket{1}_\Control$ defines the other.
}
\label{fig:ChaosPaper_CircuitPic}
\end{figure}

For simplicity, we focus on pure states
$\ket{ \psi }  \in  \Hil$.
The interferometric scheme, however,
generalizes to arbitrary
$\rho \in \mathcal{D} ( \Hil )$,
the set of density operators
(trace-one linear positive-semidefinite operators) defined on $\Hil$.
The OTOC has the form
$F_t  =  
\langle \psi | W^\dag_t  V^\dag  W_t  V  | \psi  \rangle$.
Figure~\ref{fig:ChaosPaper_CircuitPic} illustrates
the interferometric protocol.
The system-and-control composite $\Sys \Control$
ends a perfect trial in the state
\begin{align}
   \label{eq:Final_state}
   \ket{ \Psi' }  & :=  \frac{1}{ \sqrt{2} }
   [ V W_t  \ket{ \psi }  \otimes  \ket{ 0 }
     +  W_t V \ket{ \psi }  \otimes  \ket{1} ]  \, .
\end{align}

%
%
\subsection{Imperfect Hamiltonian evolution
in the interferometric scheme}
\label{section:Interf_ErrHam}

The forward and/or reverse evolution
might be implemented imperfectly:
Some unitary $U_1  =  e^{ - i H_1 t }$ might be implemented
instead of $U$,
and $U_2^\dag  =  e^{ i H_2 t }$ might be implemented
instead of $U^\dag$.
The Hamiltonians $H_1$ and $H_2$
may differ slightly from the ideal $H$.
As a result, $H_2$ might not equal $- H_1$.
The reverse evolution would not ``undo''
the forward evolution: $U_2^\dag U_1  \neq  \id$.

Multiple sources can corrupt the evolution,
including imperfect control of analog tuning.
Consider attempting to negate the Hamiltonian
by turning a knob,
which determines the angle through which a qubit is rotated,
from $\theta$ to $- \theta$.
The knob might be turned slightly past the $- \theta$ point.
Zhu \emph{et al.} mitigate analog errors with
a ``quantum clock'' in~\cite{Zhu_16_Measurement}.
Their Hamiltonian's sign depends on the state of a control qubit $\Controll$.
If $\Controll$ occupies the state $\ket{ 1 }$,
$\Sys$ evolves under $U$.
If $\Controll$ occupies $\ket{ 0 }$, $\Sys$ evolves under $U^\dag$.
A magnitude-$\pi$ rotation flips $\Controll$.
The renormalization scheme (i) mitigates the error independently and
(ii) eliminates error incurred by
depolarization of the control qubit $\Controll$ (Sec.~\ref{section:Decoher_Analytics2}).

Renormalization mitigates also errors
that threaten both the analog and quantum-clock protocols.
Each spin may experience
a small, random external magnetic field.
Additionally, the coupling strengths may vary randomly.

\subsection{Derivation of renormalization scheme
for interferometer measurements}
\label{section:Interf_PseudoDerivn}

Suppose that $\Sys \Control$ evolves imperfectly.
The joint system ends
not in the state $\ket{ \Psi' }$ [Eq.~\eqref{eq:Final_state}],
but in
\begin{align}
   \ket{ \Psi_{12}' }  =  \frac{1}{ \sqrt{2} }
   ( V U_2^\dag W U_1 \ket{ \psi }  \otimes  \ket{0}
   +  U_2^\dag  W U_1  V  \ket{\psi}  \otimes  \ket{1} )  \, .
\end{align}
By measuring the control's $\sigma^x$,
one can reconstruct
\begin{align}
   \expval{ X_\Control }  =  \Re \LParen \FInt{t} ( V , W  ) \RParen  \, ,
\end{align}
wherein
\begin{align}
   \label{eq:Define_FInt}
   \FInt{t} ( V , W  )  & :=  \expval{
   U_1^\dag W^\dag U_2 V^\dag U_2^\dag W U_1 V }
\end{align}
approximates $F_t$.
The superscript ``int'' signals that $\FInt{t} ( V , W  )$ is inferred from
the interferometric protocol.

Consider ``shielding'' each $W$
from its imperfect-unitary neighbors
by inserting identities $\id  =  U U^\dag$:
\begin{align}
   \FInt{t} ( V , W  )  & =
   \langle U_1^\dag ( U U^\dag )  W^\dag ( U U^\dag )
   U_2 V^\dag
   \nonumber \\ & \qquad \times
   U_2^\dag ( U U^\dag ) W ( U U^\dag )
   U_1 V \rangle  \, .
\end{align}
Regrouping the unitaries,
and recalling that $W_t  =  U^\dag W U$, yields
\begin{align}
   \label{eq:FInt_Help1}
   \FInt{t} ( V , W  )  & =  \langle
   ( U_1^\dag U)  W^\dag_t
   (U^\dag U_2 )  V^\dag ( U_2^\dag U )
   \nonumber \\ & \qquad \times
   W_t  (U^\dag U_1 )  V \rangle  \, .
\end{align}
Let us define a ``perturbed $V$'' through
\begin{align}
   \label{eq:Vp_Defn}
   \VInt^\dagger  :=  (U^\dag U_2 )  V^\dag ( U_2^\dag U )  \, .
\end{align}
We insert a $V^\dag V$,
formed from unperturbed unitaries,
beside the perturbed $\VInt^\dag$ in Eq.~\eqref{eq:FInt_Help1}:
\begin{align}
   \label{eq:FInt_Help2}
   \FInt{t} ( V , W  )  & =  \langle
   ( U_1^\dag U)  W^\dag_t
   V^\dag  (V \VInt^\dag )
   \nonumber \\ & \qquad \times
   W_t  (U^\dag U_1 )  V \rangle  \, .
\end{align}

Suppose that we could eliminate the
$( U_1^\dag U)$, $(V \VInt^\dag )$, and $(U^\dag U_1 )$.
$\FInt{t} ( V , W  )$ would reduce to $F_t$.
We will ``divide out'' the undesirable factors, loosely speaking.

Consider setting $W$ to $\id$,
then repeating the interferometry protocol.
This deformed protocol should require
less control than the ordinary protocol.
One would infer
\begin{align}
   \label{eq:FInt_Help3}
   \FInt{t} ( \id , V )
   =  \expval{ ( U_1^\dag U )  \VInt^\dag  ( U^\dag U_1 ) V }  \, .
\end{align}
This expectation value is of
the undesirable factors, rearranged, in Eq.~\eqref{eq:FInt_Help2}.
Hence dividing~\eqref{eq:FInt_Help2} by~\eqref{eq:FInt_Help3}
is expected to approximate the OTOC:
\begin{align}
   \label{eq:FInt_Conjecture}  \boxed{
   F_t  \approx
   \frac{ \FInt{t} ( W , V ) }{ \FInt{t} ( \id , V ) }  }  \, .
\end{align}

The approximation is expected to be strong when
the denominator is sizable:
Dividing by a number close to zero
would lead to an instability.
$\FInt{t} ( W , V )$ remains close to zero starting after the scrambling time, $t_*$
(defined as the time at which the OTOC begins
to deviate significantly from unity).
Hence Eq.~\eqref{eq:FInt_Conjecture} is expected to hold
until approximately $t = t_*$,
and the scrambling time can be inferred
from renormalized data.

Equation~\eqref{eq:FInt_Conjecture} is a conjecture
that we have motivated analytically.
Numerical support appears in Sec.~\ref{section:Interf_Numerics};
and an analytic calculation for a holographic model,
in Sec.~\ref{section:Holographic}.
Appendix~\ref{section:TInfty} motivates~\eqref{eq:FInt_Conjecture}
alternatively with an infinite-temperature limit.

Another motivating limit consists of the trivial OTOC.
Consider setting $W = V = \id$.
Every function in Eq.~\eqref{eq:FInt_Conjecture}
reduces to one.
The left-hand side equals the right-hand side
in this simple case.

\subsection{Numerical simulations of the interferometer}
\label{section:Interf_Numerics}

We consider a model of $n$ qubits with power-law decaying Ising interactions in a one-dimensional chain with open boundary conditions---the \emph{power-law quantum Ising model}.
The model's Hamiltonian is
\begin{equation}
H_\PQIM = - \sum_{\ell=1}^{\ell_0} \sum_{r=1}^{n-\ell} \frac{J}{\ell^\zeta} \sigma^z_r \sigma^z_{r+\ell} - \sum_r h^x \sigma^x_r -\sum_r   h^z_r  \sigma^z_r,
\end{equation}
wherein $J$ sets the interaction-energy scale, $\zeta$ and $\ell_0$ control the interaction range, $h^x$ denotes the transverse field, and $h^z_r$ denotes a position-dependent longitudinal field.

Most of the numerical data shown below correspond to $n=14$, $J=1$, $\zeta=6$, $\ell_0=5$, $h^x  =  1.05$, and $h^z_r=.375(-1)^r$. The OTOC operators are chosen to be $V=\sigma^x_1$ and $W=\sigma^x_n$. The renormalization scheme's power does not depend on these parameter choices. But this combination is illustrative,
causing OTOCs to grow approximately exponentially
at early times.
Simple exponential growth has proven rare
in many researchers' numerical studies of small, local spin chains.

One might expect the power-law quantum Ising model to be realizable with
immediate- and near-term quantum many-body platforms.
Possible examples include
the Rydberg-atom ensemble in~\cite{Bernien_17_Probing}.
A similar Hamiltonian has been considered independently in~\cite{Chen_17_SubsystemDiff}.

The system's initial state is taken to be either the all-$(+y)$ state or a state drawn randomly from the Hilbert space. The $+y$ state is a simple product state in the energy spectrum's center. The random state mimics the maximally mixed state's physics.
Mixed states are inconvenient to study with
the sparse-matrix techniques employed in these numerics;
random pure states serve as proxies.
Similar results can be obtained from
other initial states,
including states away from the energy spectrum's center.

The imperfect interferometric scheme is defined as follows. Starting from $H_\PQIM$, we define
the forward Hamiltonian $H_1$ and the backward Hamiltonian $H_2$. These are related to $H_\PQIM$ by the addition of random time-independent perturbations, including nearest-neighbor $\sigma^z \sigma^z$ couplings and onsite $\sigma^z$ and $\sigma^x$ fields, all of strength $\varepsilon$:
\begin{align} \label{eq:ImperfectH}
   H_1& - H_\PQIM = \nonumber \\
   & \varepsilon \sum_{r=1}^{n-1} \eta^{(1)}_{zz,r} \sigma^z_{r} \sigma^z_{r+1}
   + \varepsilon    \sum_{r=1}^{n} \eta^{(1)}_{x,r} \sigma^x_r
   +\varepsilon \sum_{r=1}^{n} \eta^{(1)}_{z,r}    \sigma^z_r  \, ,
\end{align}
and
\begin{align}
H_2& - H_\PQIM = \nonumber \\
& \varepsilon \sum_{r=1}^{n-1} \eta^{(2)}_{zz,r} \sigma^z_{r} \sigma^z_{r+1} + \varepsilon \sum_{r=1}^{n} \eta^{(2)}_{x,r} \sigma^x_r +\varepsilon \sum_{r=1}^{n} \eta^{(2)}_{z,r} \sigma^z_r.
\end{align}
Each of $\eta^{(i)}_{zz,r}$, $\eta^{(i)}_{z,r}$, and $\eta^{(i)}_{x,r}$ is a random variable drawn uniformly from $\left[ - \frac{1}{2} ,  \,  \frac{1}{2}  \right]$.
Each run involves one instance of $H_1$ and one instance of $H_2$.
Each plot shows the OTOC's real part, unless otherwise stated.
All times are measured in units in which the nearest-neighbor coupling
$J = 1$.

\begin{figure}[hbt]
\centering
\includegraphics[width=.48\textwidth, clip=true]{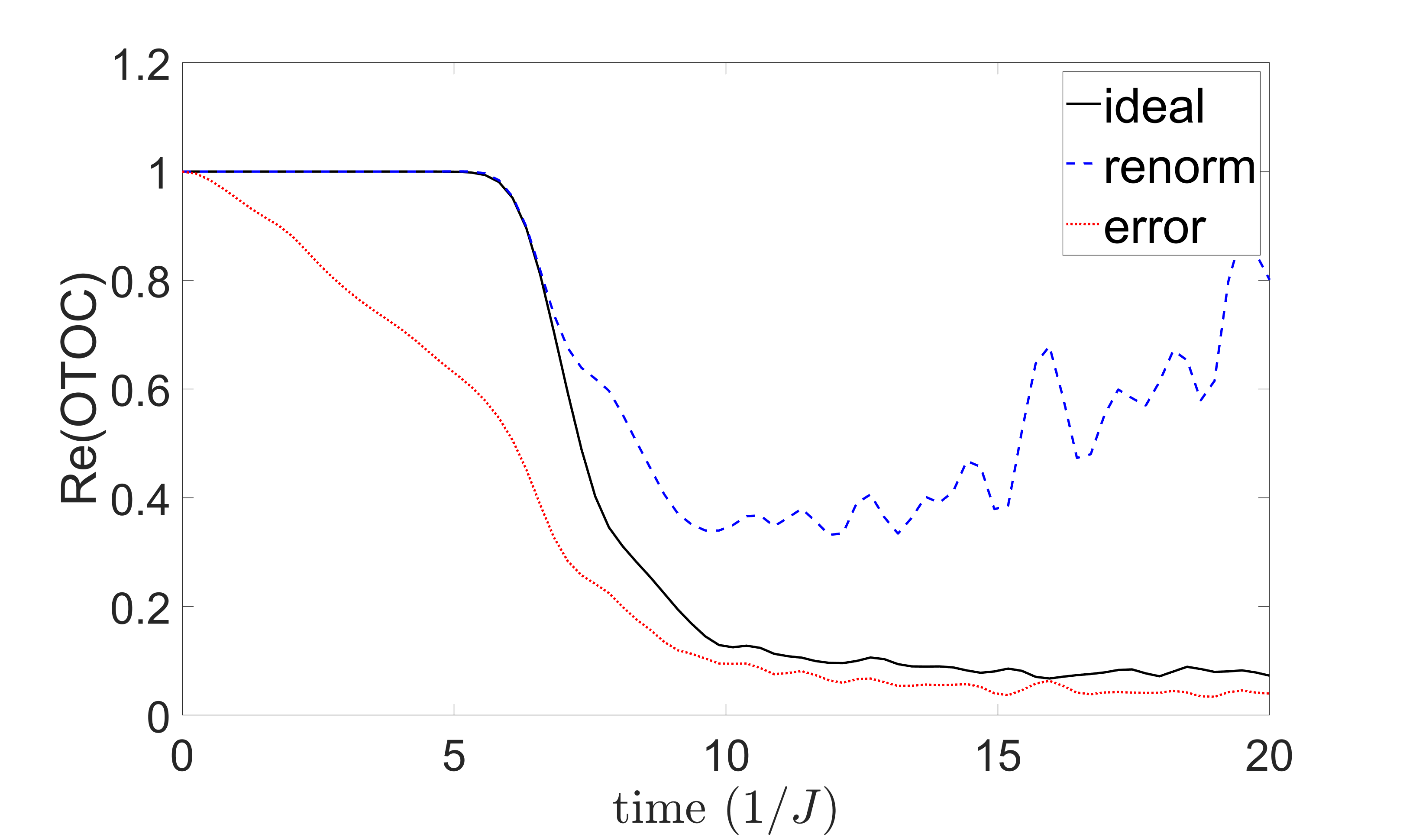}
\caption{\caphead{Interferometric renormalization results:}
Single run of the power-law quantum Ising model with $n=14$ spins, initial state all $+y$, and error $\varepsilon =.2$.
The three curves correspond to the ideal OTOC (black), the imperfect value (red, dotted), and the renormalized result obtained from Eq.~\eqref{eq:FInt_Conjecture} (blue, dashed). The imperfect value indicates an incorrect scrambling time. But the renormalized value remains close to the ideal up to the true scrambling time.}
\label{fig:n14eps0p2}
\end{figure}

\begin{figure}[hbt]
\centering
\includegraphics[width=.48\textwidth, clip=true]{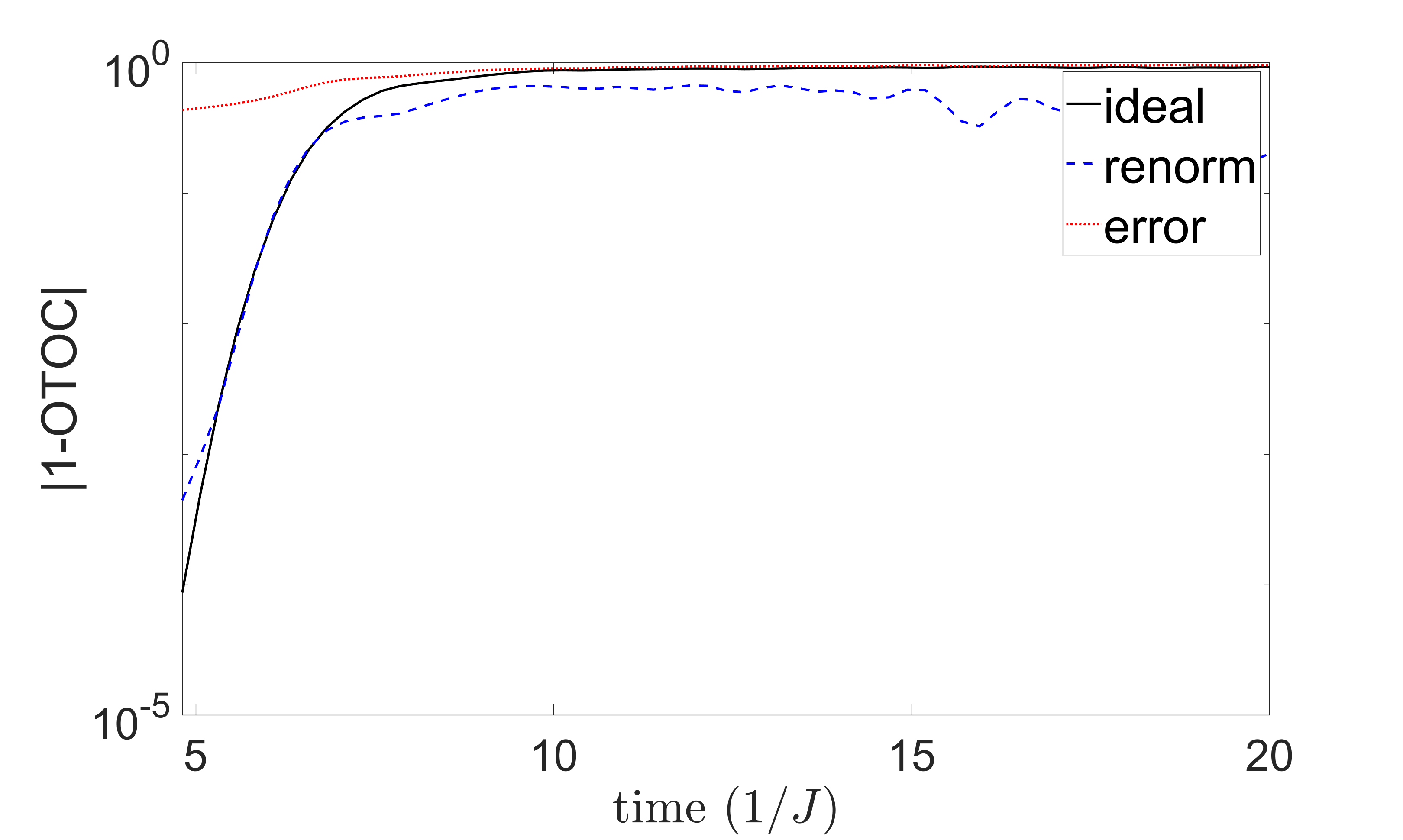}
\caption{\caphead{Interferometric renormalization results:} Same data as in Figure~\ref{fig:n14eps0p2}, on a semilogarithmic plot. The ideal OTOC's early-time exponential growth is visible, although this behavior is unusual for a small spin chain. The ideal value (black) is compared again with the imperfect value (red, dotted) and the renormalized value (blue, dashed). Remarkably, the renormalized value's exponential growth rate is very close to the ideal value over more than three decades. In fact, this behavior persists over several more decades at earlier times (not shown).}
\label{fig:n14eps0p2_log}
\end{figure}

Figures~\ref{fig:n14eps0p2} and \ref{fig:n14eps0p2_log} show
the results of one run of the renormalization scheme for
$n=14$ spins with $\varepsilon=.2$
and the all-$(+y)$ initial state. This choice of $\varepsilon$ corresponds to imperfections that are $\pm 10\%$ of the nearest-neighbor coupling, a quite sizable perturbation. Nevertheless, while the imperfect signal deviates substantially from the ideal result, the renormalized value remains close to the ideal up to scrambling time.

\begin{figure}[hbt]
\centering
\includegraphics[width=.48\textwidth, clip=true]{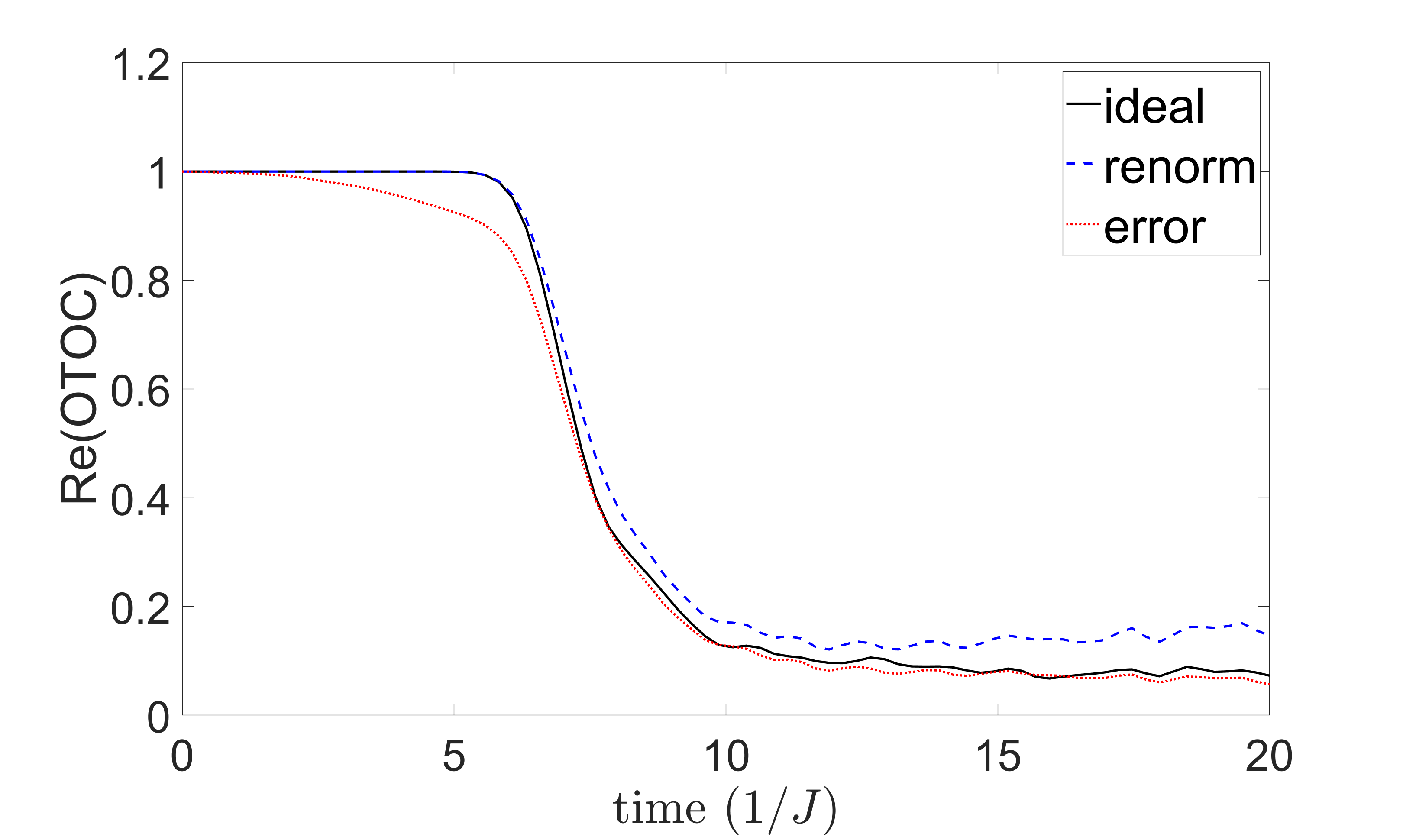}
\caption{\caphead{Interferometric renormalization results:}
Single run of the power-law quantum Ising model with $n=14$ spins, initial state all $+y$, and error $\varepsilon =.1$. The three curves correspond to the ideal OTOC (black), the imperfect value (red, dotted), and the renormalized result obtained from Eq.~\eqref{eq:FInt_Conjecture} (blue, dashed).}
\label{fig:n14eps0p1}
\end{figure}

\begin{figure}[hbt]
\centering
\includegraphics[width=.48\textwidth, clip=true]{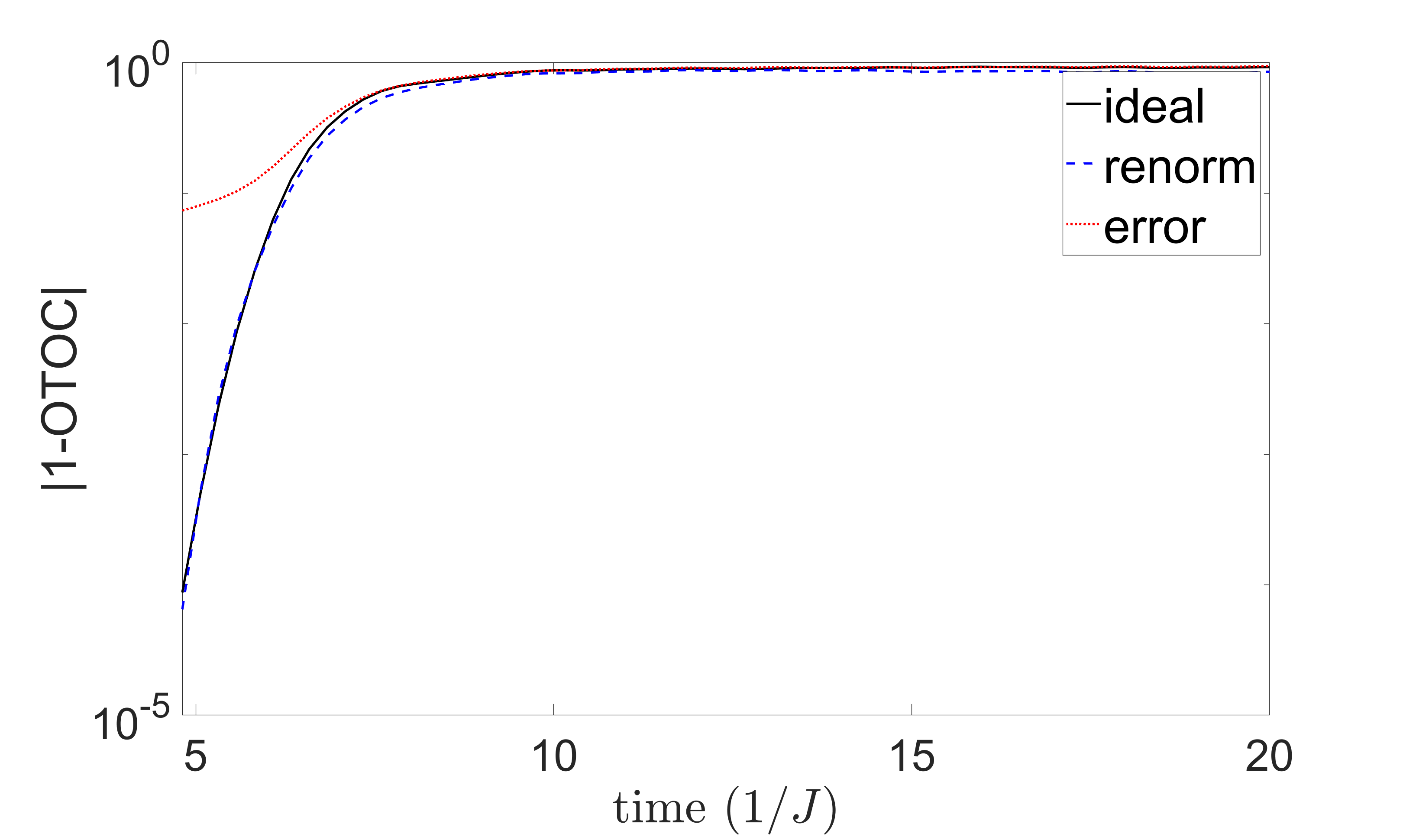}
\caption{\caphead{Interferometric renormalization results:} Same data as in Figure~\ref{fig:n14eps0p1}, on a semilogarithmic plot.}
\label{fig:n14eps0p1_log}
\end{figure}

\begin{figure}[hbt]
\centering
\includegraphics[width=.48\textwidth, clip=true]{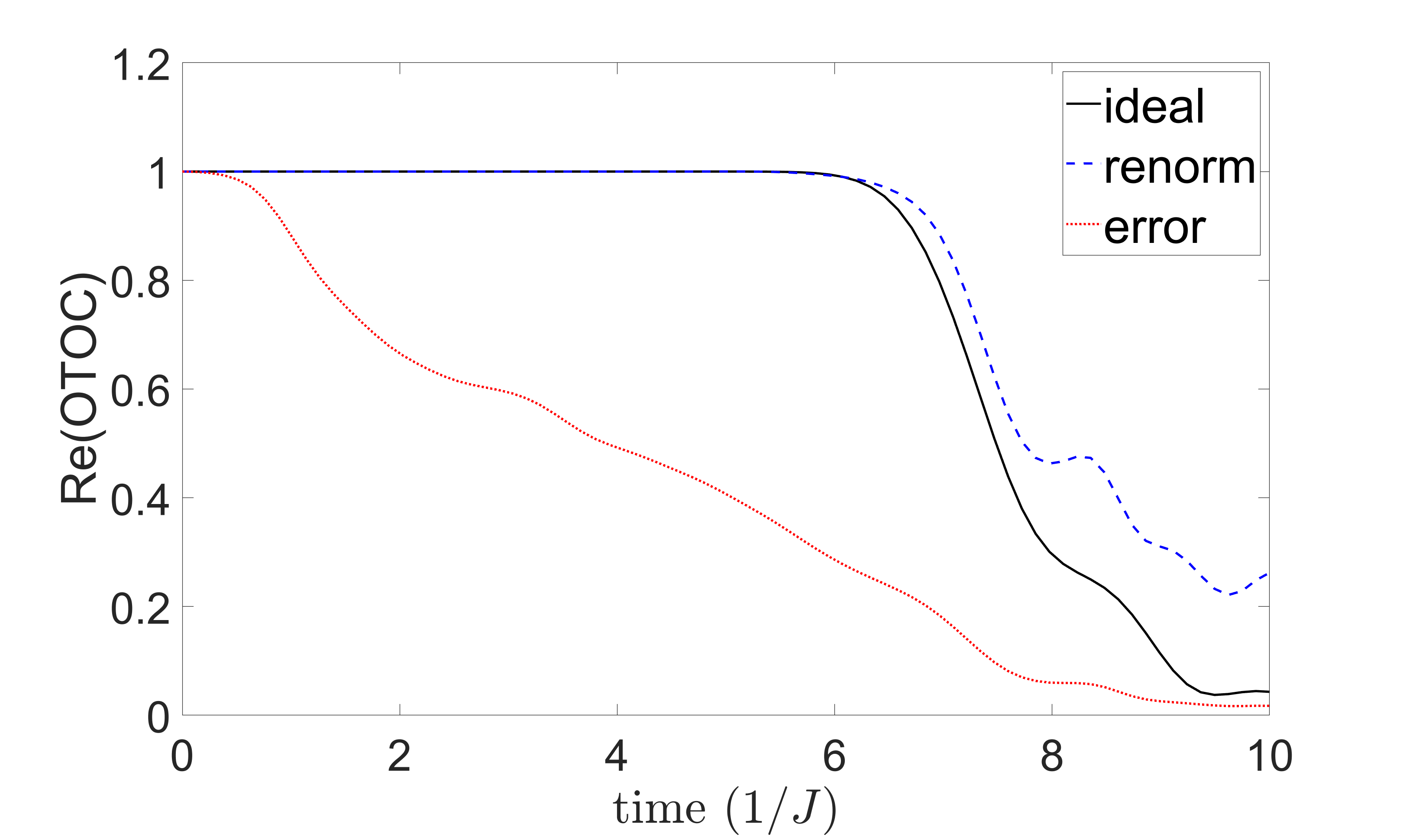}
\caption{\caphead{Interferometric renormalization results:}
Single run of the power-law quantum Ising model with $n=14$ spins, initial state all $+y$, and error $\varepsilon =.3$. The three curves correspond to the ideal OTOC (black), the imperfect value (red, dotted), and the renormalized result obtained from Eq.~\eqref{eq:FInt_Conjecture} (blue, dashed).}
\label{fig:n14eps0p3}
\end{figure}

\begin{figure}[hbt]
\centering
\includegraphics[width=.48\textwidth, clip=true]{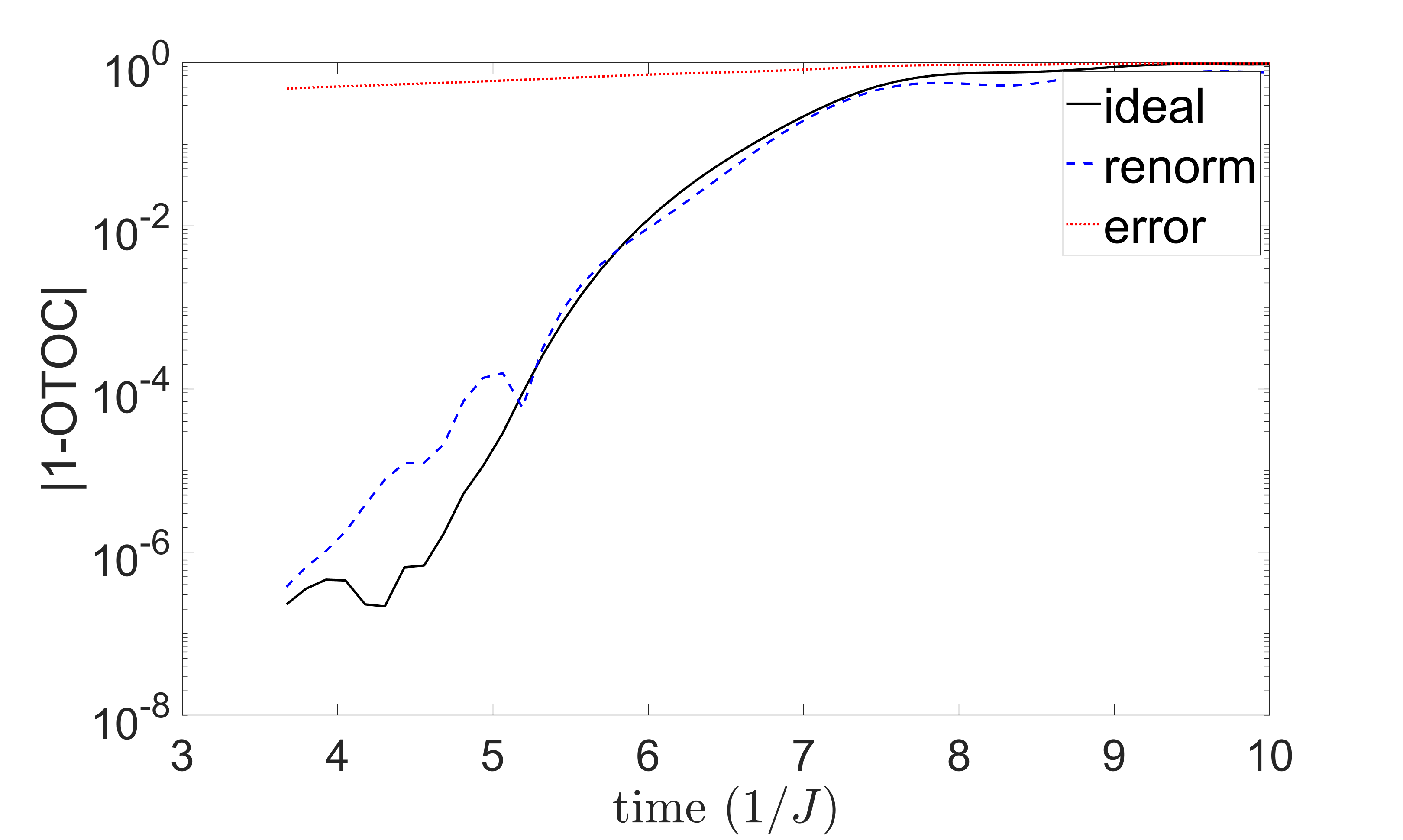}
\caption{\caphead{Interferometric renormalization results:}
Same data as in Figure~\ref{fig:n14eps0p3}, on a semilogarithmic plot.
The curves jag because
the sign of $1-F_t$ varies and the time grid is coarse.
The value of $1  -  F_t$ passes through zero as it changes sign.
Hence a semilogarithmic plot of $|1-F_t|$ spikes downward dramatically.
This early-time region can be studied
 with a finer time grid, to resolve these jags.
 But observing such small values of $1-F_t$
in near-term experiments is impractical.
Hence we omitted a finer-grained study.}
\label{fig:n14eps0p3_log}
\end{figure}

Figures~\ref{fig:n14eps0p1} and \ref{fig:n14eps0p1_log} show the results of one run with $\varepsilon$ reduced to $\varepsilon=.1$. Now, the agreement between the ideal and the renormalized values is remarkable at early times. Yet the two values still diverge somewhat after the scrambling time.
Outside the regime in which
the renormalization is expected to approximate $F$,
i.e., after $t_*$, the imperfect value tracks the ideal OTOC better than
the renormalized value does.
We can also push the results in the opposite direction, considering $\varepsilon=.3$, as shown in Figures~\ref{fig:n14eps0p3} and \ref{fig:n14eps0p3_log}. Clearly, the renormalized value's quality decreases as $\varepsilon$ increases.
But, even here, the early-time agreement is reasonable.

\begin{figure}[hbt]
\centering
\includegraphics[width=.48\textwidth, clip=true]{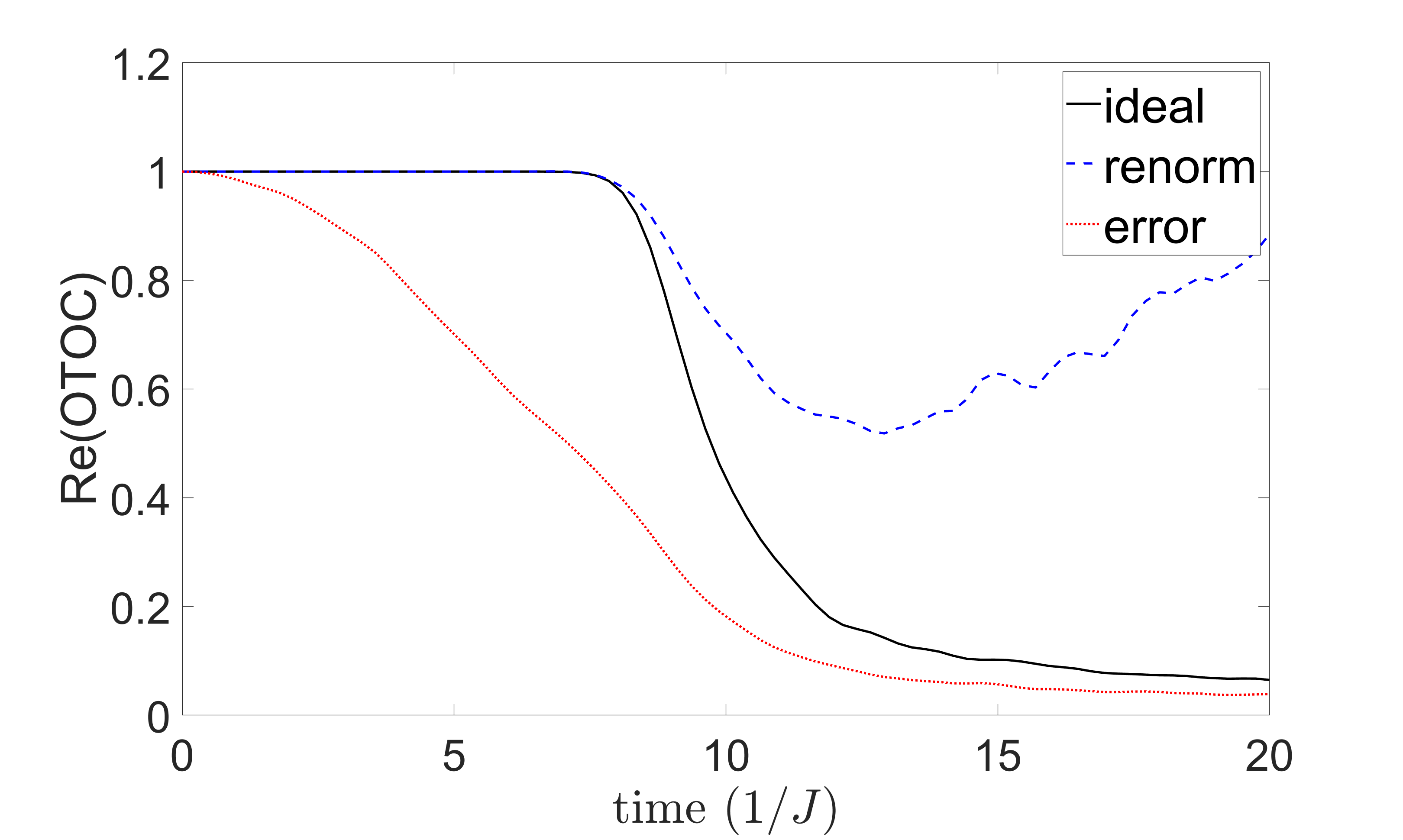}
\caption{\caphead{Interferometric renormalization results:}
Single run of the power-law quantum Ising model with $n=18$ spins, initial state all $+y$, and error $\varepsilon =.2$. The three curves correspond to the ideal OTOC (black), the imperfect value (red, dotted), and the renormalized result obtained from Eq.~\eqref{eq:FInt_Conjecture} (blue, dashed).}
\label{fig:n18eps0p2}
\end{figure}

\begin{figure}[hbt]
\centering
\includegraphics[width=.48\textwidth, clip=true]{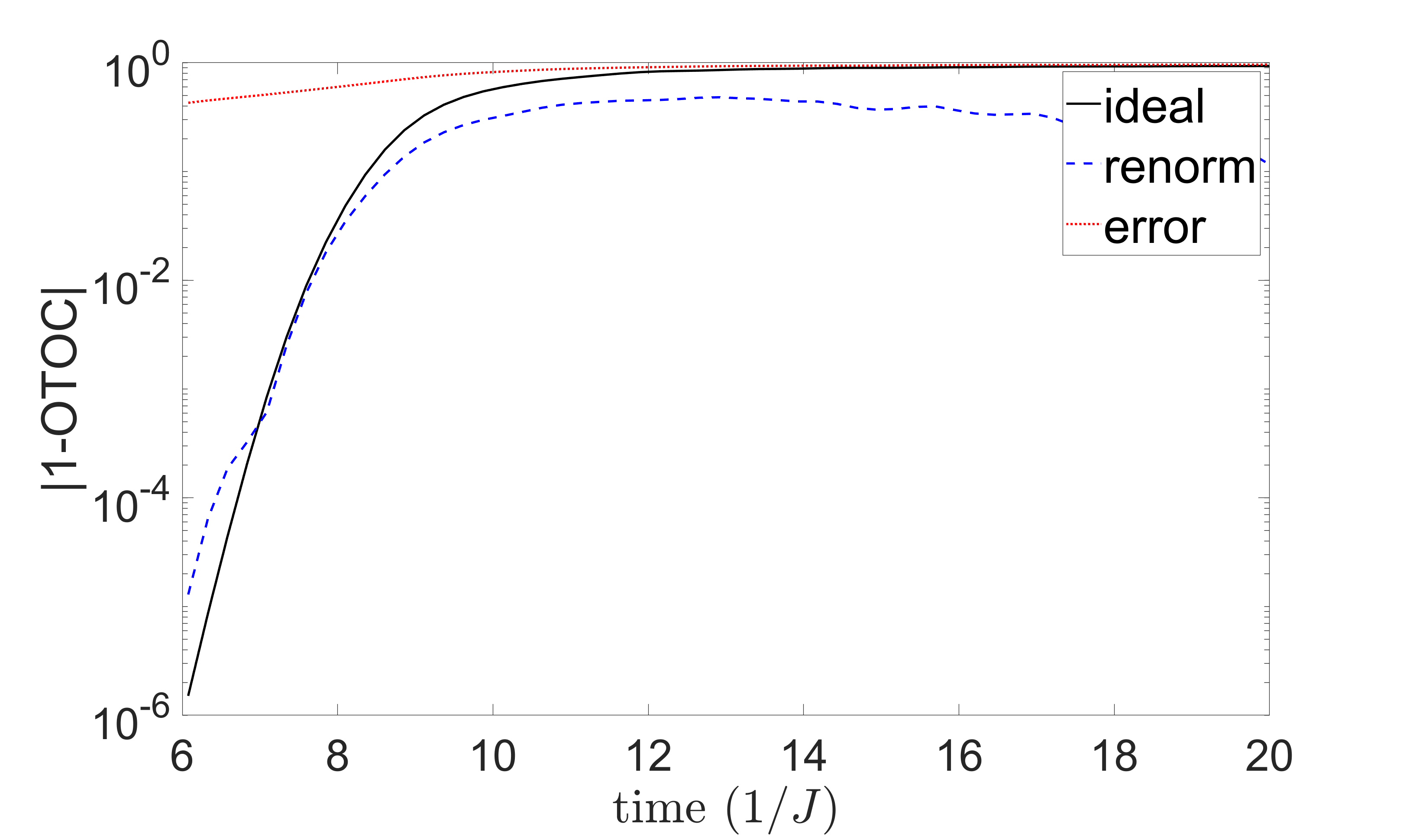}
\caption{\caphead{Interferometric renormalization results:}
Same data as in Figure~\ref{fig:n14eps0p1}, on a semilogarithmic plot. }
\label{fig:n18eps0p2_log}
\end{figure}

We can also check the system-size dependence. Substantially increasing the system size to $n=18$, with $\varepsilon=.2$, leads to Figures~\ref{fig:n18eps0p2} and \ref{fig:n18eps0p2_log}. The quality of the early-time match between the ideal and renormalized values is of comparable quality to the $n=14$ quality. But the time scale at which the two deviate is noticeably earlier, though still around the scrambling time.

\begin{figure}[hbt]
\centering
\includegraphics[width=.48\textwidth, clip=true]{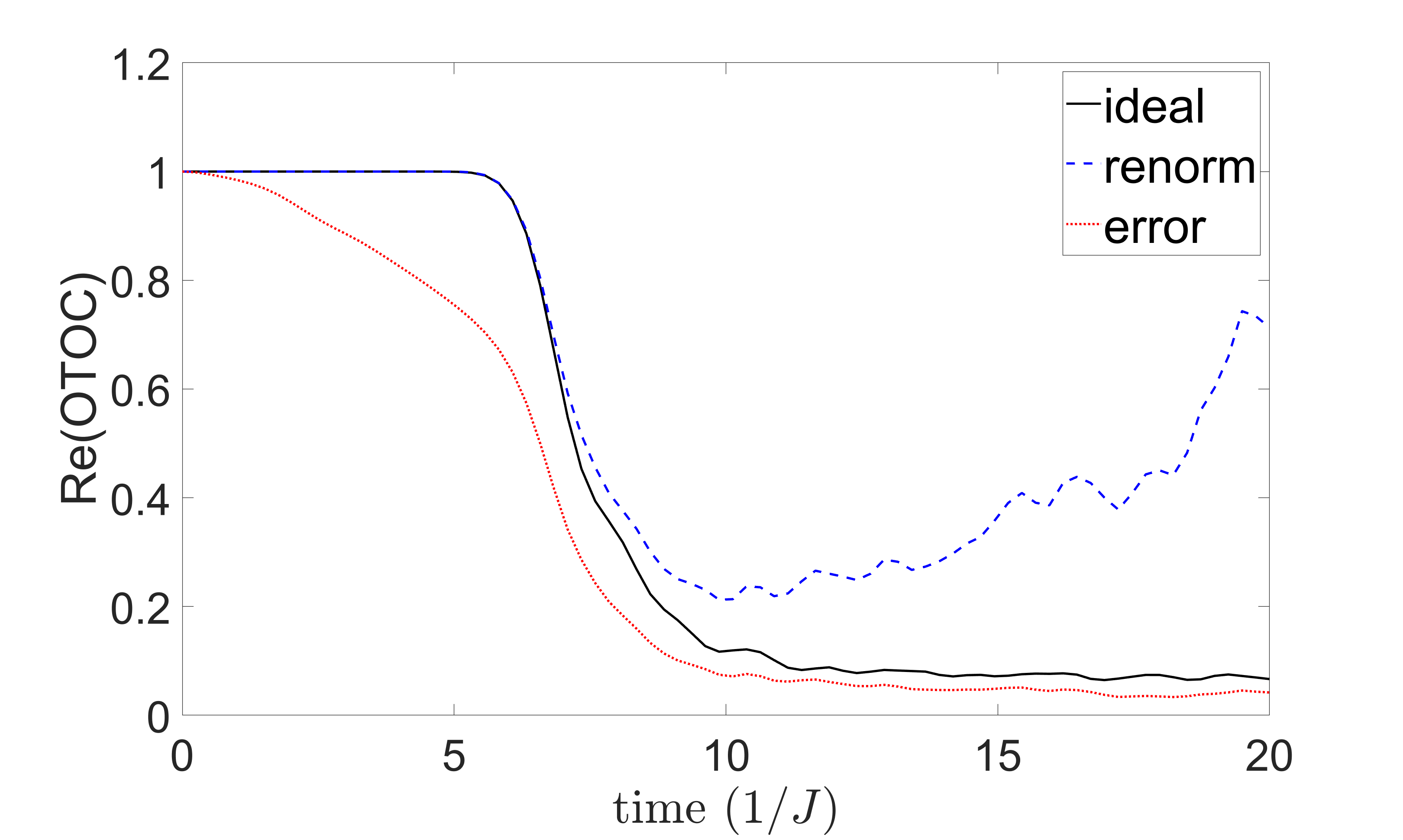}
\caption{\caphead{Interferometric renormalization results:}
Single run of the power-law quantum Ising model with $n=14$ spins, a random initial state, and error $\varepsilon =.2$. The three curves correspond to the ideal OTOC (black), the imperfect value (red, dotted), and the renormalized result obtained from Eq.~\eqref{eq:FInt_Conjecture} (blue, dashed).}
\label{fig:n14eps0p2_rand}
\end{figure}

\begin{figure}[hbt]
\centering
\includegraphics[width=.48\textwidth, clip=true]{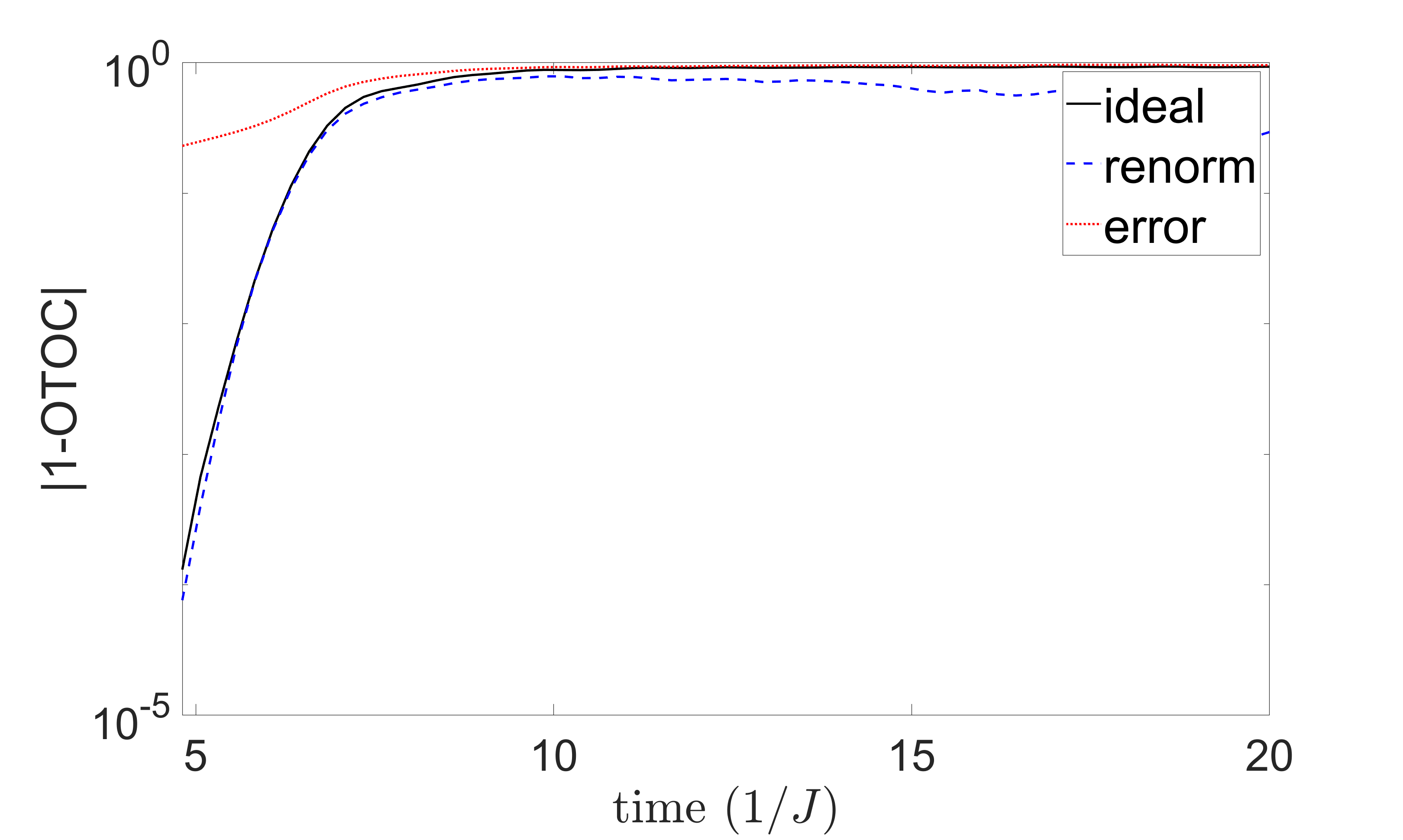}
\caption{\caphead{Interferometric renormalization results:}
Same data as in Figure~\ref{fig:n14eps0p2_rand}, on a semilogarithmic plot. }
\label{fig:n14eps0p2_rand_log}
\end{figure}

The renormalized value's quality depends also on the initial state. For example, if we choose a random initial state, the renormalized value matches the ideal result better. Such a random state mimics a maximally mixed state.
Hence the renormalization scheme could work best with the infinite-temperature state.
This likelihood is promising for nuclear-magnetic-resonance (NMR) experiments,
whose initial states tend to be highly mixed~\cite{Li_16_Measuring,Wei_16_NuclearSpinOTOC}.
Numerical results for $n=14$ spins and a random initial state are shown in Figures~\ref{fig:n14eps0p2_rand} and \ref{fig:n14eps0p2_rand_log}. As claimed, the agreement between the renormalized and ideal values is enhanced relative to the all-$(+y)$ initial state.

\section{Example \#2: Weak measurement}
\label{section:Weak}

Weak measurements can be used to infer
the OTOC experimentally.
A \emph{weak measurement} barely disturbs
the measured system.
Refraining from damaging the quantum state
is often desirable but comes with a tradeoff:
A weak measurement extracts little information.
But averaging over weak-measurement trials
reproduces strong-measurement statistics.
Also, weak measurements offer experimental access to
OTOCs and to more-fundamental \emph{quasiprobabilities}~\cite{YungerHalpern_17_Jarzynski,NYH_17_Quasi}.

The weak-measurement protocol for inferring the OTOC is detailed in
Appendix A of~\cite{YungerHalpern_17_Jarzynski} and
is simplified in~\cite[Sec.~II]{NYH_17_Quasi}.\footnote{
Let $\Sites$ denote the number of degrees of freedom,
e.g., the number of spins in a chain.
In the original protocol, each measured observable $O$
equals a product of $\Sites$ local operators $O_j$:
$O  =  \otimes_{ j = 1}^\Sites  O_j$.
In the simplified protocol, each observable
nontrivially transforms just one spin.
}
We focus on the simplified protocol, though
the renormalization scheme is expected to extend
to the original protocol.

\begin{figure}[hbt]
\centering
\includegraphics[width=.45\textwidth, clip=true]{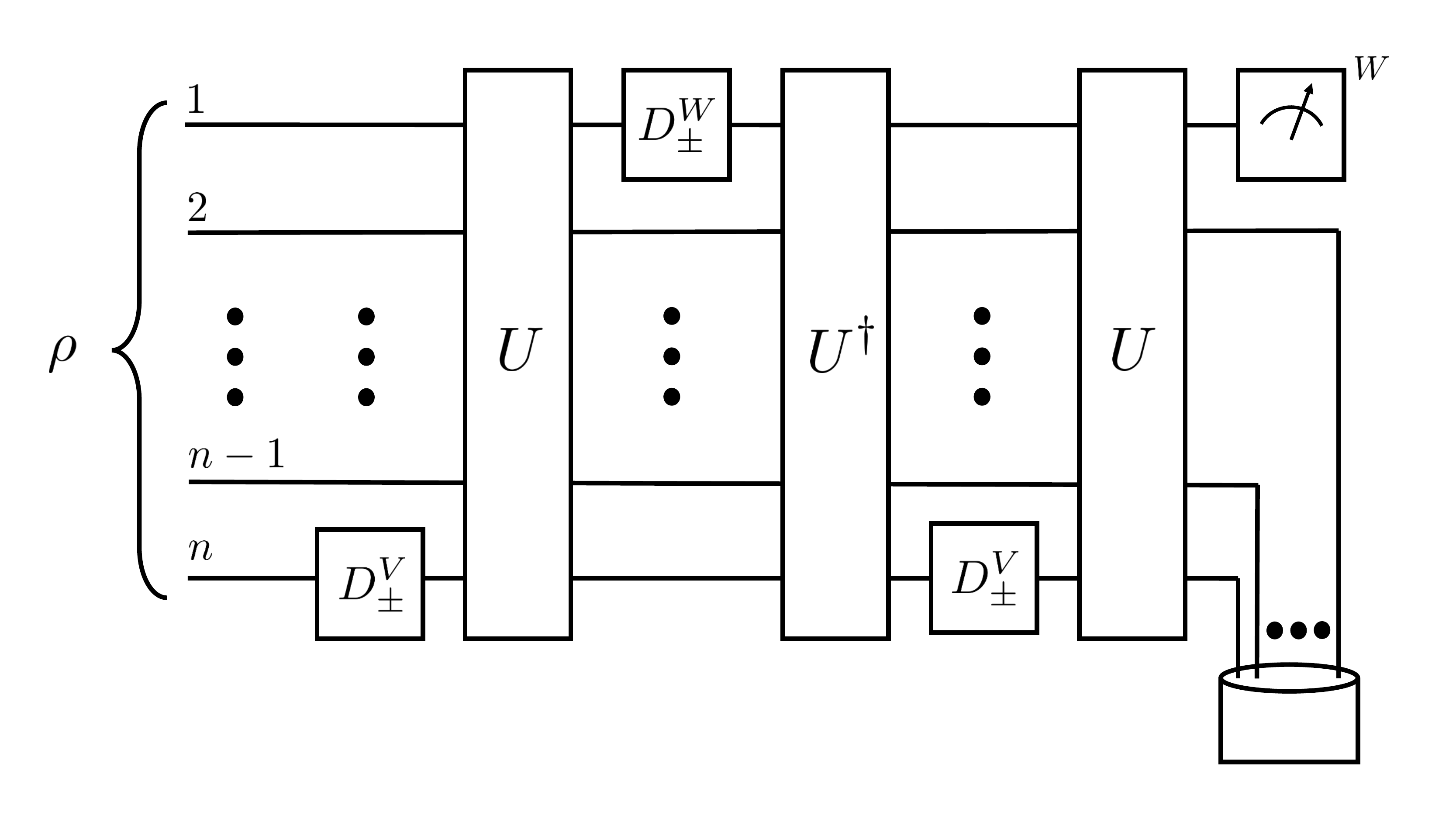}
\caption{\caphead{Weak-measurement protocol for measuring
the out-of-time-ordered correlator (OTOC):}
This figure was adapted from Figure~3b of~\cite{NYH_17_Quasi}.
The protocol is illustrated with
a quantum circuit for a chain of $\Sites$ spins.
The system is prepared in an arbitrary state $\rho$.
$V$ and $W$ represent local observables.
(The protocol extends to non-Hermitian unitaries $V$ and $W$.)
Each box labeled $D_\pm^{ V }$ represented, in~\cite{NYH_17_Quasi},
a weak measurement of
a projector $\ProjV{v_\ell}$ onto
the eigenvalue-$v_\ell$ eigenspace of
the observable $V$.
Here, the boxes represent weak measurements of $V$,
e.g., Pauli operators.
The $D_\pm^W$ boxes serve analogously.
The intrinsic system Hamiltonian $H$ generates
the time-evolution operator $U$.
Two forward evolutions $U$, and one reverse evolution $U^\dag$,
alternate with three weak measurements
and one strong $W$ measurement.}
\label{fig:Full_circuit}
\end{figure}

Figure~\ref{fig:Full_circuit} reviews
the weak-measurement protocol.
Hamiltonian errors are modeled,
and the renormalization approximation is derived,
in Sec.~\ref{section:Weak_PseudoDerivn}.
Numerical simulations in Sec.~\ref{section:Weak_Numerics}
support the scheme.

%
%
\subsection{Derivation of renormalization scheme
for weak-measurement data}
\label{section:Weak_PseudoDerivn}

The weak-measurement circuit contains
a forward evolution $U$,
followed by a reverse evolution $U^\dag$,
followed by another $U$.
Each evolution might be implemented imperfectly.
We denote the implemented unitaries by
$U_1  :=  e^{ - i H_1 t }$,
$U_2^\dag  :=  e^{ i H_2 t }$, and
$U_3  :=  e^{ - i H_3 t }$.
The erroneous Hamiltonians
$H_1 , H_2 , H_3  \neq  H$.

From many imperfect weak-measurement trials,
one can infer the approximation
\begin{align}
   \label{eq:Quasi_err_help2}
   \SumKDWeak{\rho} ( v_1, w_2, v_2, w_3 )
   &  :=  \Tr ( U_1^\dag U_2 U_3^\dag \ProjW{ w_3 }  U_3
   \ProjV{v_2}
   \nonumber \\ & \qquad \quad \times
   U_2^\dag  \ProjW{w_2}  U_1
   \ProjV{v_1} \rho  )
\end{align}
to the OTOC.
Equation~\eqref{eq:Quasi_err_help2}
follows from Eq.~(37) of~\cite{NYH_17_Quasi}.
More generally,
\begin{align}
   \label{eq:Quasi_err_help3}
   \FWeak{t} ( A, B, C, D )  :=  \Tr  \left(
   U_1^\dag U_2  U_3^\dag  A^\dag  U_3
   B^\dag
   U_2^\dag  C  U_1
   D  \rho  \right)  \, .
\end{align}

Consider ``shielding'' each $W$ from
its imperfect-unitary neighbors
with factors of $\id = U U^\dag$.
We regroup unitaries,
then recall $W_t  :=  U^\dag W U$:
\begin{align}
   \label{eq:Quasi_err_help4}
   \FWeak{t} ( W , V , W , V )  & =  \Tr  \LParen
   [ U_1^\dag U_2  U_3^\dag U ]  W^\dag_t
   [ U^\dag  U_3  V^\dag   U_2^\dag  U ]
   \nonumber \\ & \qquad \quad \times
   W_t
   [ U^\dag U_1  V ]  \rho  \RParen  \, .
\end{align}

We would almost recover the OTOC if we could replace
the $U^\dag  U_3^\dag  V^\dag  U_2^\dag  U$ with $V^\dag$
and the $U^\dag  U_1  V$ with $V$.
Let us ape the replacement.
We insert an $\id  =  V V^\dag$ rightward of
the $U^\dag  U_3^\dag  V^\dag  U_2^\dag  U$
and one leftward of the $U^\dag  U_1  V$.
Regrouping unitaries yields
\begin{align}
   \FWeak{t} ( W , V , W , V ) 
   \label{eq:Quasi_err_help6}
   & =   \Tr  \LParen  [ U_1^\dag U_2  U_3^\dag U ]   \W^\dag_t
            [  U^\dag  U_3  V^\dag  U_2^\dag  U  V  ]
            \nonumber \\ & \qquad  \times
           V^\dag   \W_t   V
        [  V^\dag U^\dag  U_1  V  ]   \rho  \RParen \, .
\end{align}

Equation~\eqref{eq:Quasi_err_help6} would equal the OTOC
if the bracketed factors were removed.
One might expect the bracketed factors to have roughly the size
\begin{align}
   & \Tr  \left( [ U_1^\dag U_2  U_3^\dag U ]
         [  U^\dag  U_3  V^\dag  U_2^\dag  U  V  ]
         [  V^\dag  U^\dag  U_1  V  ]   \rho  \right)
    \nonumber \\   & \qquad =
    \Tr  \left(  U_1^\dag  U_2  V^\dag  U_2^\dag   U_1  V   \rho  \right)
    \\ & \label{eq:Pillows} \qquad =
    \FWeak{t} ( \id , V, \id, V )  \, .
\end{align}

We wish to remove the bracketed factors' influence
on $\FWeak{t} ( W, V, W, V )$.
One might attempt to do so by
dividing~\eqref{eq:Quasi_err_help6} by~\eqref{eq:Pillows}:
\begin{align}
   \label{eq:Weak_conjecture0}
   F_t  \approx
   \frac{  \FWeak{t} ( W , V , W , V )  }{
              \FWeak{t} ( \id , V, \id, V ) }  \, .
\end{align}
But consider setting $V$ to $\id$.
The left-hand side reduces to one.
So does the right-hand side's denominator.
But the numerator evaluates to
\begin{align}
   \label{eq:Quasi_err_help7}
   & \Tr \LParen U_1^\dag U_2 U_3^\dag
   W^\dag  U_3 U_2^\dag
   W U_1 \rho  \RParen  \\
   & = \FWeak{t} ( W, \id , W , \id )  \, .
\end{align}
Hence we divide the right-hand side of Eq.~\eqref{eq:Weak_conjecture0}
by~\eqref{eq:Quasi_err_help7}:
\begin{align}
   \label{eq:Weak_conjecture}  \boxed{
   F_t  \approx
   \frac{  \FWeak{t} ( W , V , W , V )  }{
              \FWeak{t} ( \id , V, \id, V )  \,
              \FWeak{t} ( W, \id , W , \id ) } }  \, .
\end{align}


The weak-measurement conjecture~\eqref{eq:Weak_conjecture}
requires a $W$-dependent factor.
The interferometer conjecture~\eqref{eq:FInt_Conjecture} does not.
Why, physically?

The Hamiltonian is negated only once in the interferometry protocol.
Hence equating $V$ with $\id$ in Eq.~\eqref{eq:Define_FInt} enables
the $U_2$ to cancel the $U_2^\dag$.
That cancellation frees the $W^\dag$ to cancel the $W$.
Hence $\FInt{t}(V, W)$ reduces to one if $V = \id$,
regardless of what $W$ equals.

In contrast, the Hamiltonian is negated twice in
the weak-measurement protocol.
$U_3$ can fail to equal $U_2$.
Hence the $U_3$ in Eq.~\eqref{eq:Quasi_err_help7}
can fail to cancel the $U_2^\dag$,
despite $V$'s equaling $\id$.
Hence the $W^\dag$ cannot ``reach'' the $W$ to cancel it.
A $W$-dependent factor must be divided out in~\eqref{eq:Weak_conjecture}.

\subsection{Numerical simulations of the weak-measurement scheme}
\label{section:Weak_Numerics}

We numerically study the weak-measurement renormalization scheme
in Eq.~\eqref{eq:Weak_conjecture}.
For simplicity, we restrict to chaotic parameters of the power-law quantum Ising model.
Various other limits give similar results, however.
All the plots below are for a system size of $n=12$.
This choice is merely numerically convenient:
Larger sizes requires sparse-matrix techniques,
and the weak-measurement scheme requires
simulations of three time evolutions.
(In contrast, the interferometric scheme requires that
only two time evolutions be simulated.)

\begin{figure}[hbt]
\centering
\includegraphics[width=.48\textwidth, clip=true]{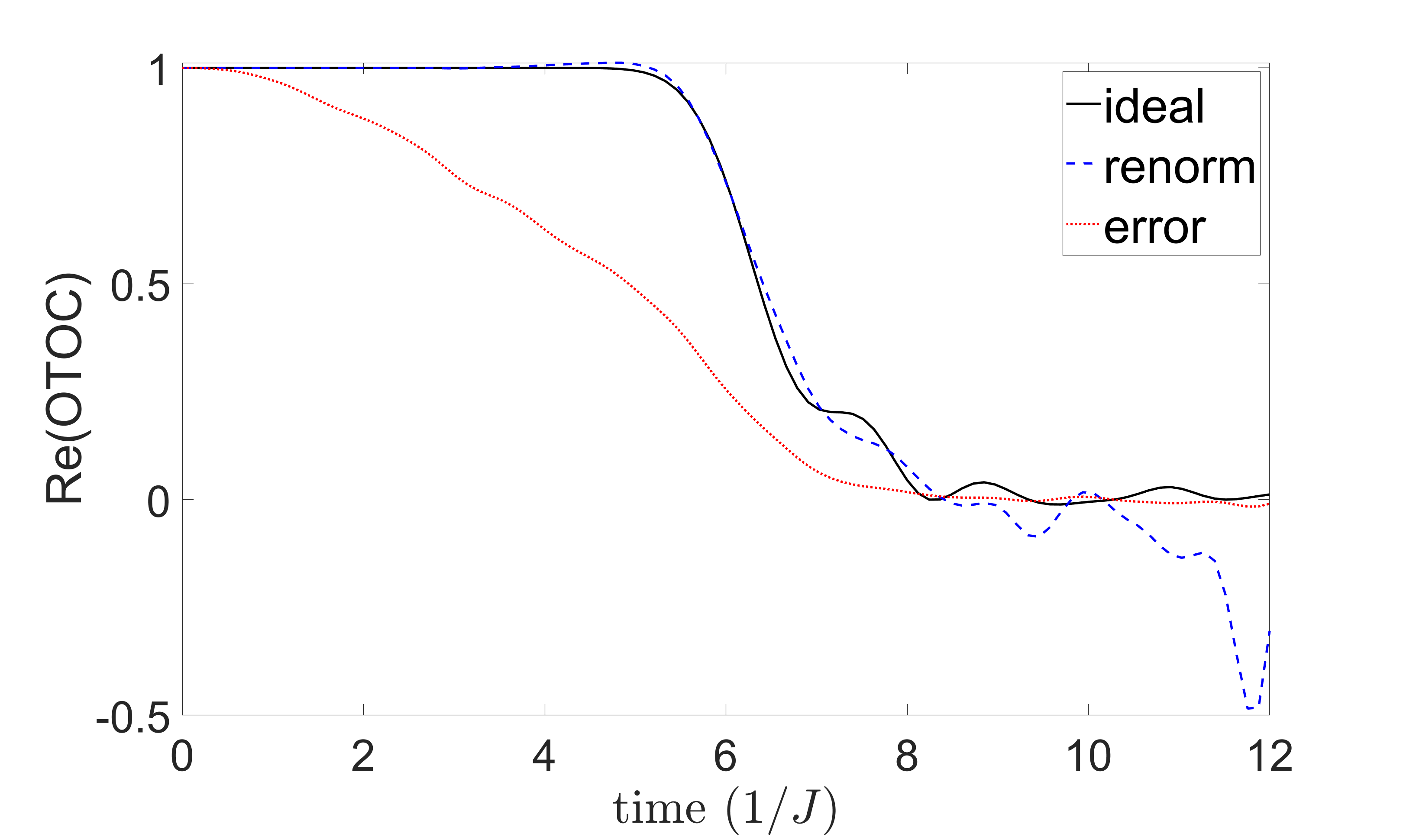}
\caption{\caphead{Weak-measurement renormalization:}
Power-law quantum Ising model with $n=12$ spins, initial state all $+y$, and error $\varepsilon=.2$,
with the weak-measurement renormalization protocol~\eqref{eq:Weak_conjecture}.}
\label{fig:n12eps0p2_weak}
\end{figure}

\begin{figure}[hbt]
\centering
\includegraphics[width=.48\textwidth, clip=true]{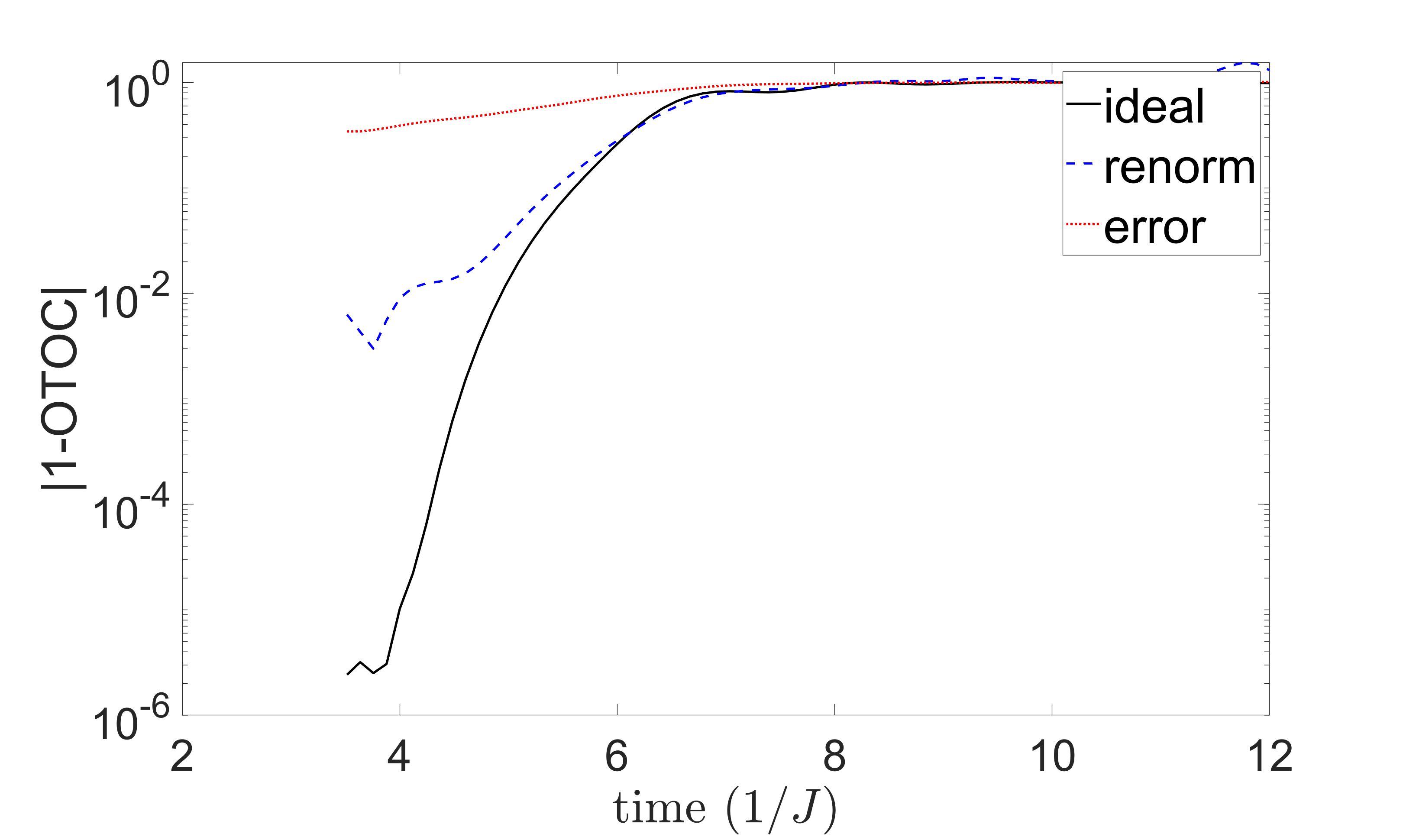}
\caption{\caphead{Weak-measurement renormalization:}
Same data as in Figure~\ref{fig:n12eps0p2_weak}, on a semilogarithmic plot. }
\label{fig:n12eps0p2_weak_log}
\end{figure}

Figures~\ref{fig:n12eps0p2_weak} and \ref{fig:n12eps0p2_weak_log}
compare the ideal, imperfect, and renormalized values of
a weak measurement of the OTOC.
Each of $U_1$, $U_2$, and $U_3$ is generated by
a Hamiltonian that differs from the ideal by an amount $\varepsilon=.2$.
(See Eq.~\eqref{eq:ImperfectH} and the surrounding discussion.)
Even for this large value of $\varepsilon$,
and though the weak-measurement scheme involves three imperfect time evolutions
(instead of only two),
the early-time agreement between the ideal and renormalized values remains reasonably good.

\begin{figure}[hbt]
\centering
\includegraphics[width=.48\textwidth, clip=true]{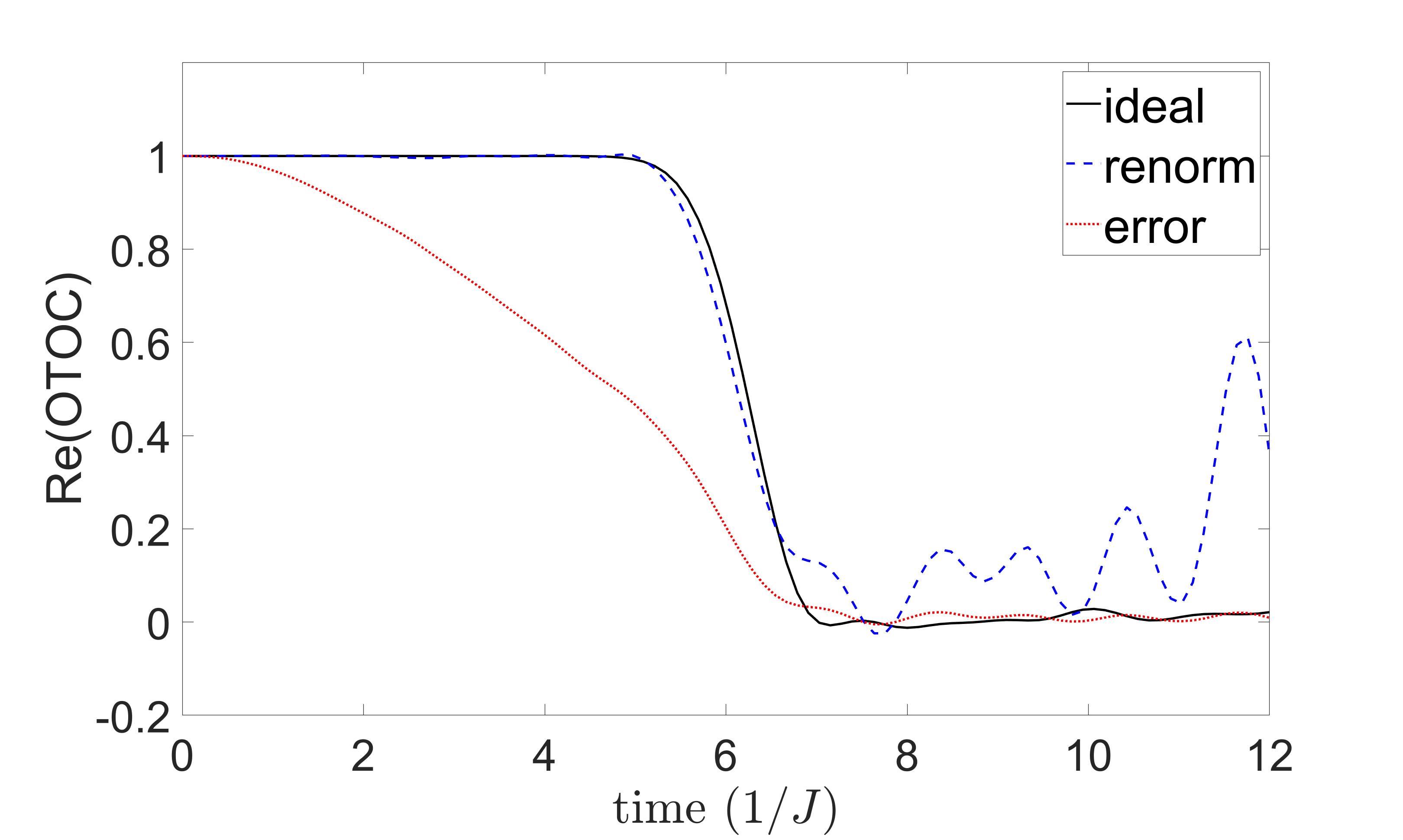}
\caption{\caphead{Weak-measurement renormalization:}
Power-law quantum Ising model with $n=12$ spins, a random initial state, and error $\varepsilon=.2$,
with the weak-measurement renormalization protocol~\eqref{eq:Weak_conjecture}.}
\label{fig:n12eps0p2_weak_rand}
\end{figure}

\begin{figure}[hbt]
\centering
\includegraphics[width=.48\textwidth, clip=true]{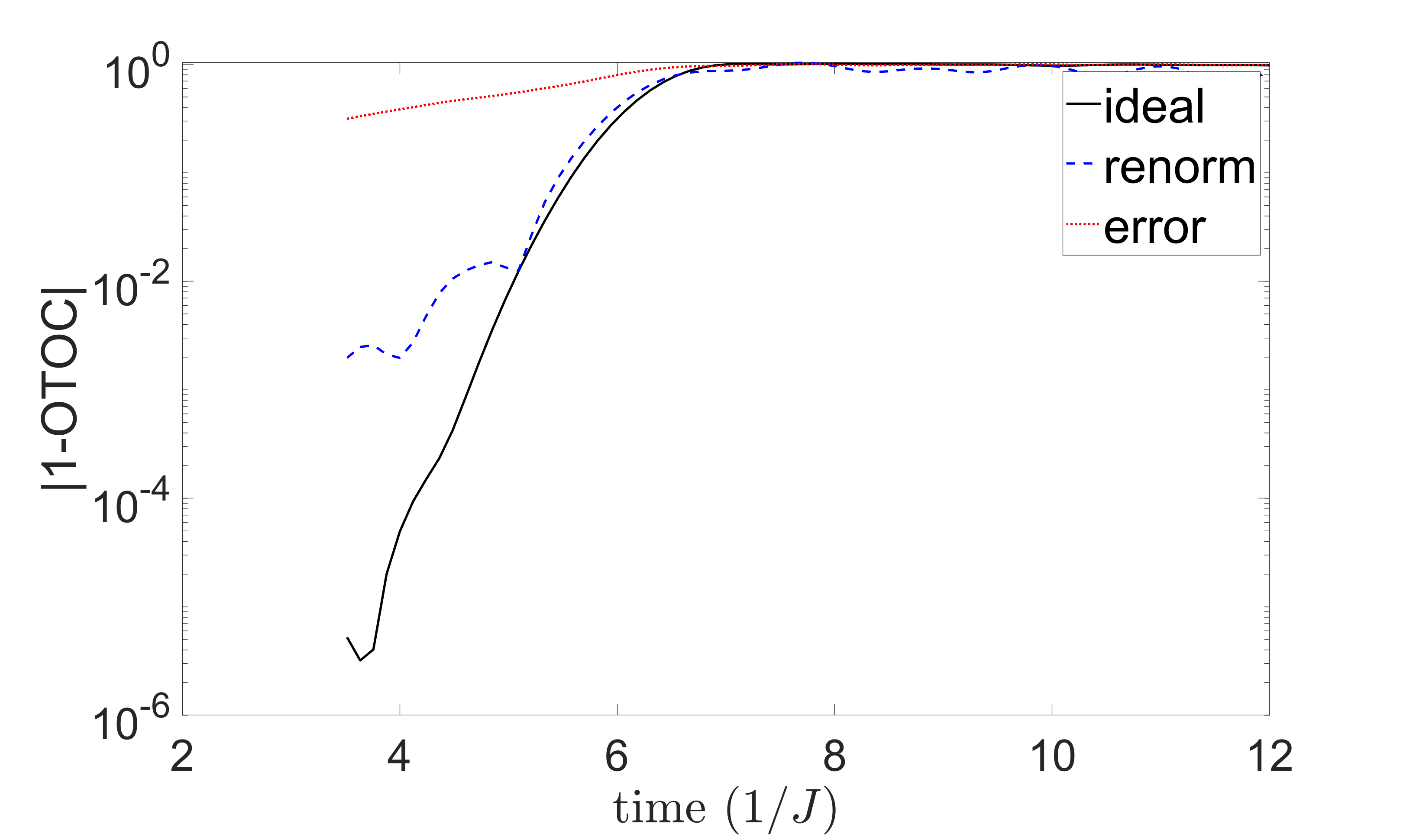}
\caption{\caphead{Weak-measurement renormalization:}
Same data as in Figure~\ref{fig:n12eps0p2_weak_rand}, on a semilogarithmic plot. }
\label{fig:n12eps0p2_weak_rand_log}
\end{figure}

Figures~\ref{fig:n12eps0p2_weak_rand} and \ref{fig:n12eps0p2_weak_rand_log} show the same situation, except with a random initial state, instead of an all $+y$ initial state. As with the interferometric renormalization scheme, the random state leads to improved agreement at early times and a longer period of agreement at later times.

\begin{figure}[hbt]
\centering
\includegraphics[width=.48\textwidth, clip=true]{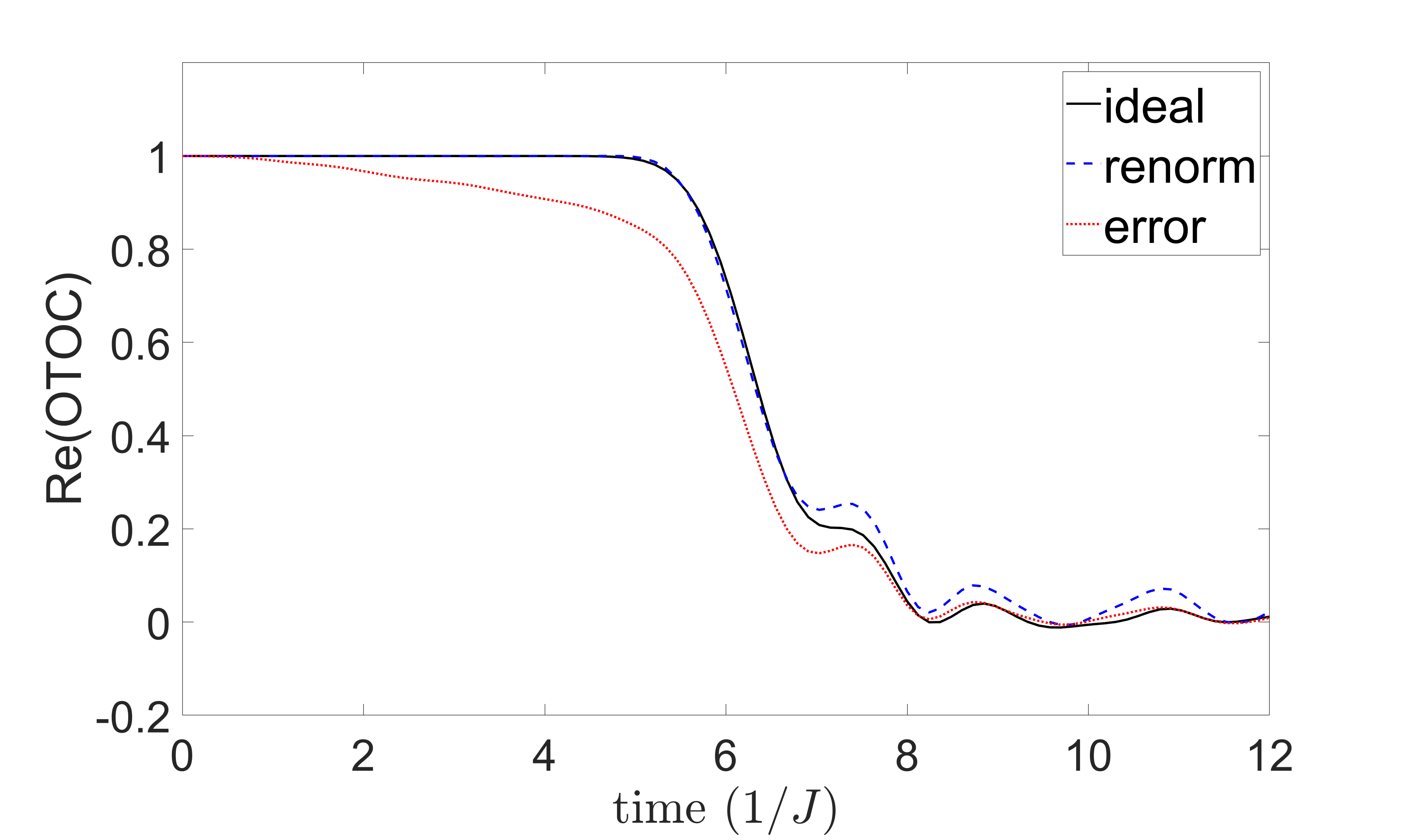}
\caption{\caphead{Weak-measurement renormalization:}
Power-law quantum Ising model with $n=12$ spins, initial state all $+y$, and error $\varepsilon=.1$, with the weak-measurement renormalization protocol~\eqref{eq:Weak_conjecture}.}
\label{fig:n12eps0p1_weak}
\end{figure}

\begin{figure}[hbt]
\centering
\includegraphics[width=.48\textwidth, clip=true]{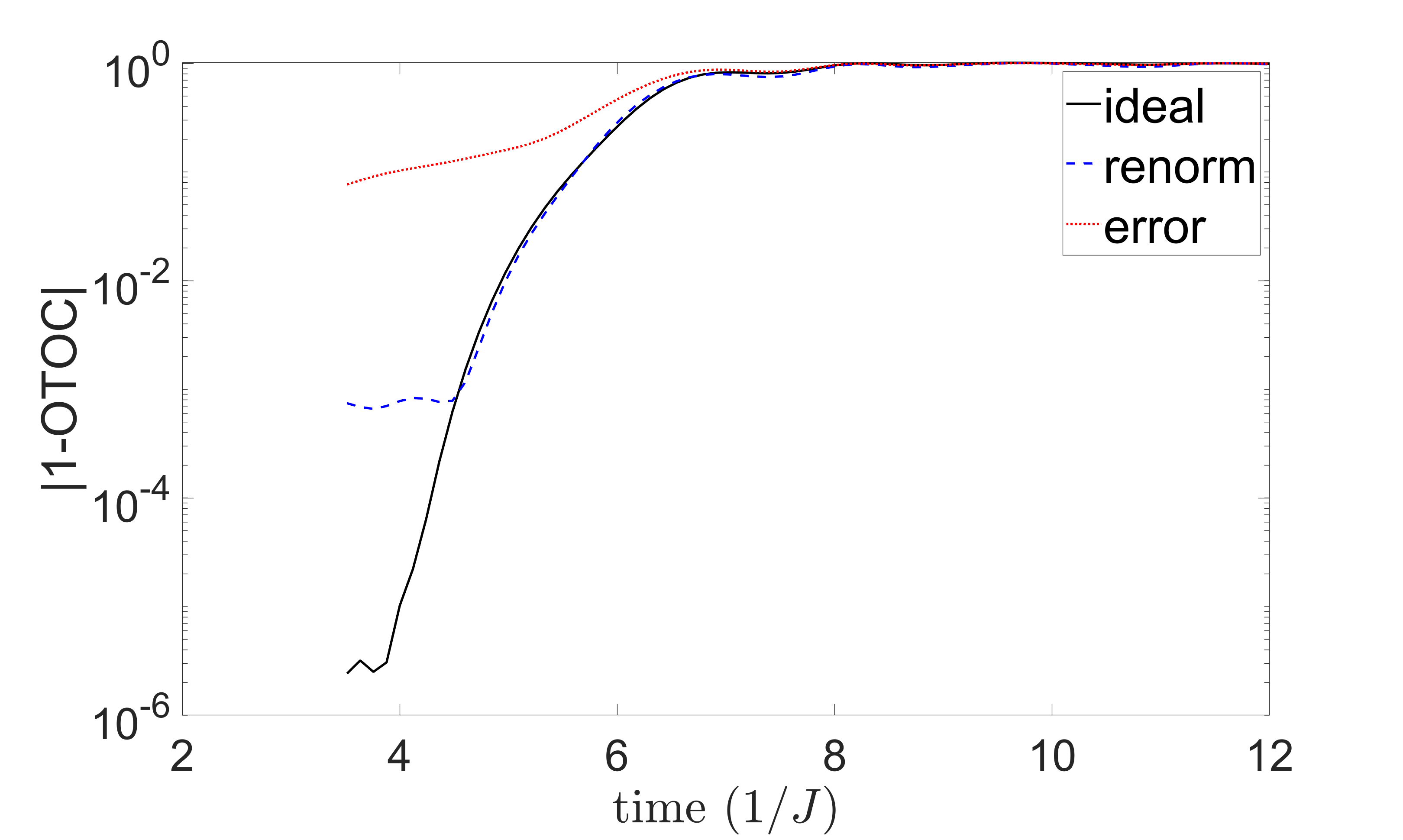}
\caption{\caphead{Weak-measurement renormalization:}
Same data as in Figure~\ref{fig:n12eps0p1_weak}, on a semilogarithmic plot. }
\label{fig:n12eps0p1_weak_log}
\end{figure}

Figures~\ref{fig:n12eps0p1_weak} and \ref{fig:n12eps0p1_weak_log} show
the weak-measurement renormalization scheme with $\varepsilon =.1$.
Downsizing the error improves
the agreement between
the ideal and renormalized signals.
There is some disagreement at very early times.
But the signal there is already so small,
we expect it to be difficult to access with near-term experiments.

\section{Decoherence by the environment}
\label{section:Decoher}

Sections~\ref{section:Interferometer} and~\ref{section:Weak} detailed
how to infer about $F_t$ from
protocols marred by Hamiltonian errors.
Unitaries modeled the evolutions.
But the environment can couple to the system~\cite{Syzranov_17_Out,Gonzalez_Alonso_18_Resilience,Zhang_18_Information}.
The state can evolve under
a nonunitary channel $\mathcal{E}$~\cite{NielsenC10}.
Nevertheless, we show, renormalization facilitates
the recovery of $F_t$.

$F_t$ can be recovered perfectly despite
two instances of decoherence.
First, Garttner \emph{et al.} have measured an OTOC of
over 100 trapped ions~\cite{Garttner_16_Measuring}.
We generalize their measurement scheme
in Sec.~\ref{section:Decoher_Analytics}.
We then suppose that
the ions' state depolarizes probabilistically.
Renormalization enables the retrieval of $F_t$,
an analytical proof shows,
without channel tomography.

Second, we return to the interferometric measurement
of Sec.~\ref{section:Interferometer}.
We suppose that the control qubit suffers
probabilistic decoherence.
Again, renormalization enables the inference of $F_t$
without channel tomography.

Section~\ref{section:Decoher_Numerics} complements
the analytics with numerics.
The power-law quantum Ising model is coupled to another spin chain.
The interaction and environmental Hamiltonians
remain unchanged as the system Hamiltonian is reversed.

\subsection{Exact recovery of $F_t$ despite
probabilistic depolarization of the system
during a generalization of the ion-trap protocol}
\label{section:Decoher_Analytics}

The ion-trap experiment in~\cite{Garttner_16_Measuring}
motivates this section.
We review their protocol in Sec.~\ref{section:Ion_review}
and generalize their set-up in Sec.~\ref{section:Set_up_ions}.
The system could decohere during each unitary evolution.
We model decoherence with probabilistic depolarization.
Section~\ref{section:Ideal_ions} concerns the ideal limit.
Section~\ref{section:Imperfect_ions} concerns the general case.
The exact value of $F_t$ can be extracted
via renormalization.
The extraction requires no channel tomography.

\subsubsection{Motivation: Ion-trap protocol}
\label{section:Ion_review}

Garttner \emph{et al.} implemented the following protocol:
\begin{enumerate}

   \item
   Prepare the ions in the eigenstate
   $\ket{ \Plus }  :=  \ket{ + }^{ \otimes \Sites }$ of
   the Pauli product $\otimes_{j = 1}^\Sites  \sigma_j^x$.

   \item
   Evolve the system forward in time under
   the all-to-all Ising Hamiltonian
   $H  =  \frac{J}{\Sites} \sum_{ i < j }
     \sigma^z_i  \sigma^z_j$.
   The coupling strength is denoted by $J$.

   \item
   Rotate the qubits counterclockwise through an angle $\phi$
   about the $x$-axis, with\footnote{
   \label{footnote:GlobalW}
   This $W$ acts nontrivially on every qubit.
   A conventional $W$, described in earlier sections,
   acts nontrivially on just a small subsystem.
   Experimental practicalities motivated the many-qubit $W$.
   But this $W$ equals a product of single-qubit operators.
   See~\cite{Garttner_16_Measuring} for further discussion.}
   $W  =  e^{ - i  \phi \sum_j  \sigma^x_j }$.

   \item
   Evolve the system backward, under $-H$.

   \item
   Measure the $i^\th$ spin's $x$-component, $V  =  \sigma^x_i$,
   for any $i  =  1 , 2 , \ldots , \Sites$.
   The value of $i$ does not matter,
   due to the system's translational invariance.
   Averaging the outcomes over trials yields
   the expectation value
   \begin{align}
      \label{eq:Garttner_OTOC}
      & \langle \Plus |
      U^\dag e^{ i  \phi \sum_j  \sigma^x_j }  U   \sigma^x_i
      U^\dag  e^{ - i  \phi \sum_j  \sigma^x_j }  U   \sigma^x_i
      |  \Plus  \rangle
      \nonumber \\ &
      =  \langle \Plus | W^\dag_t  V^\dag W_t V | \Plus \rangle  \, .
   \end{align}

\end{enumerate}

The ions could couple to the environment during either evolution.
A quantum channel $\mathcal{E}$ would
evolve the system's state~\cite{NielsenC10}.
We model the channel with probabilistic depolarization.
The environment has some probability of
mapping the state to the maximally mixed state $\id / \Dim$,
wherein $\Dim$ denotes the Hilbert space's dimensionality.

%
%
%
\subsubsection{General set-up}
\label{section:Set_up_ions}

Let $\Sys$ denote a quantum system
associated with a Hilbert space $\Hil$
of dimensionality ${\rm dim} ( \Hil )  =  \Dim$.
In~\cite{Garttner_16_Measuring},
$\Sys$ consists of $\Sites$ qubits.
Hence $\Dim = 2^\Sites$.

The natural Hamiltonian $H$ generates
the ideal evolution $U  :=  e^{ - i H t}$.
The actual evolution is imperfect:
$\Sys$ has a probability $p \in [0, 1]$
of undergoing $U$ and
a probability $1 - p$ of depolarizing totally
to $\id / \Dim$.
This probabilistic depolarization evolves a state $\sigma$ as
\begin{align}
   \label{eq:Define_depol}
   \Depol{p} ( \sigma )  =  p  \, U \sigma U^\dag
   +  ( 1 - p )  \frac{ \id }{ \Dim }  \, .
\end{align}

The reverse evolution is ideally $U^\dag$.
The actual evolution has a probability $1 - q$ of
depolarizing the state completely:
\begin{align}
   \label{eq:Define_depol_rev}
   \Depoll{q} ( \sigma )  =  q  \, U^\dag \sigma U
   +  ( 1 - q )  \frac{ \id }{ \Dim }  \, .
\end{align}
The forward and reverse probabilities
need not equal each other: $p \neq q$.
An experimentalist need not know the probabilities' values,
to infer $F_t$:
Renormalization will cancel $p$ and $q$ from the calculation.

The operators $W$ and $V$ are unitary:
$W^\dag W  =  V^\dag V  =  \id$.
Additionally, $V$ is Hermitian and traceless:
$V^\dag  =  V$, and $\Tr ( V )  =  0$.
Pauli operators satisfy these assumptions.

Let $v$ denote an arbitrary eigenvalue of $V$.
Let $\DegenV_v$ denote the set of degeneracy parameters
for the $v$ eigenspace.
$\Sys$ begins in a state $\rho$
supported just in the $v$ eigenspace:
\begin{align}
   \label{eq:V_rho}
   \rho  =  \sum_{ \DegenV_v, \DegenV_v' }
   q_{ \DegenV_v ,  \DegenV_v' }
   \ketbra{ v, \DegenV_v }{ v, \DegenV_v' }  \, .
\end{align}
The coefficients satisfy the normalization condition
$|  q_{ \DegenV_v ,  \DegenV_v' }  |^2  =  1$.

The protocol proceeds as follows:
$\Sys$ is prepared in the state $\rho$.
The system is evolved under $\Depol{p}$,
then under $W$,
then under $\Depoll{q}$.
The system ends in the state
\begin{align}
   \label{eq:RhoPrime}
   \rho'  & :=  \Depoll{q}  \LParen  W  \Depol{p} ( \rho )  W^\dag  \RParen  \\
   & =  pq \, W_t \rho W_t^\dag
   +  ( 1 - p q )  \frac{ \id }{ \Dim }  \, .
\end{align}
$V$ is measured.
This process is repeated in each of many trials.
Averaging the outcomes yields
the expectation value $\Tr ( V \rho' )$.
The renormalization scheme requires also
a set of trials in which $W  =  \id$.

\subsubsection{Ideal case}
\label{section:Ideal_ions}

Suppose that $p = q = 1$.
The system ends in the state
$\rho'_\ideal  =  W_t  \rho  W^\dag_t$.
The expectation value of $V$ becomes
\begin{align}
   \label{eq:Depol_help0}
   \Tr ( V \rho'_\ideal )  =  \Tr ( V W_t \rho W^\dag_t )
   =  \Tr \LParen  W^\dag_t  V^\dag  W_t  \rho  \RParen  \, .
\end{align}
The second equality follows from the trace's cyclicality
and the Hermiticity of $V$.
By Eq.~\eqref{eq:V_rho}, $\frac{ V }{ v }  \rho  =  \rho$.
Hence inserting a $V / v$ leftward of $\rho$ yields
\begin{align}
   \label{eq:Depol_help1}
   \frac{1}{ v }  \Tr ( V \rho'_\ideal )  =  F_t  \, .
\end{align}
The expectation value is proportional to the OTOC.

\subsubsection{Imperfect evolution and renormalization}
\label{section:Imperfect_ions}

The expectation value of $V$ becomes
\begin{align}
   \FDepol{t}{p}{q} ( W, V)
   & :=  \Tr ( V \rho' )    \\
  \label{eq:Depol_help3}
  & =  \frac{ p q }{ v }  F_t  \, .
\end{align}
This expression follows from
the tracelessness of $V$.

$W$ must equal $\id$ in another set of trials.
The expectation value of $V$ reduces to
\begin{align}
   \label{eq:Depol_help4}
   \FDepol{t}{p}{v} ( \id, V )  =  p q v  \, ,
\end{align}
by $V \rho = v \rho$ and the normalization of $\rho$.

Consider dividing the right-hand side of Eq.~\eqref{eq:Depol_help3} by
the right-hand side of Eq.~\eqref{eq:Depol_help4}.
The quotient is proportional to the OTOC:
\begin{align}
   \label{eq:Depol_result}  \boxed{
   F_t  =  v^2  \:
              \frac{ \FDepol{t}{p}{q} ( W , V ) }{
                        \FDepol{t}{p}{q} ( \id, V ) }  }  \, .
\end{align}

%
%
%


%
%
%
\subsection{Exact recovery of $F_t$ despite
probabilistic depolarization of the control qubit
in the interferometric protocol}
\label{section:Decoher_Analytics2}

The interferometric protocol
relies on a control qubit $\Control$ (Sec.~\ref{section:Interf_Setup}).
$\Control$ is prepared in the state $\ket{ + }$.
Suppose that it decoheres.
We model the decoherence with
probabilistic depolarization:
\begin{align}
   \label{eq:Decoh_int_1}
   \ketbra{+}{+}  & \mapsto
   p \ketbra{ + }{ + }  +  (1 - p )  \,  \frac{ \id }{2}  \\
   & = \frac{1}{2}
   [  \ketbra{0}{0}  +  \ketbra{1}{1}
      + p ( \ketbra{0}{1}  +  \ketbra{1}{0} ) ]   \, .
\end{align}

The joint system-and-control state $\ket{ \Psi }$
must be replaced with
\begin{align}
   \label{eq:Decoh_int_2}
   \rho_{ \Sys \Control }  & =
   \ketbra{ \psi }{ \psi }  \otimes
   \frac{1}{2}
   [  \ketbra{0}{0}  +  \ketbra{1}{1}
      + p ( \ketbra{0}{1}  +  \ketbra{1}{0} ) ]  \, .
\end{align}
The interferometer maps the joint state to
\begin{align}
   \label{eq:Decoh_int_3}
   \rho'_{ \Sys \Control }  & = \frac{1}{2}
   [ V W_t  \ketbra{ \psi }{ \psi }  W_t  V
         \otimes  \ketbra{0}{0}
     +  W_t V  \ketbra{ \psi }{ \psi }  V  W_t
         \otimes  \ketbra{1}{1}
         \nonumber \\ & \qquad
     + p (  V W_t  \ketbra{ \psi }{ \psi }  V  W_t
               \otimes  \ketbra{0}{1}
               \nonumber \\ & \qquad \qquad
              +  W_t  V  \ketbra{ \psi }{ \psi }  W_t  V
                  \otimes  \ketbra{1}{0} ) ]  \, .
\end{align}
We recast $\rho'_{ \Sys \Control }$ in terms of
the eigenstates $\ket{ + }$ and $\ket{ - }$ of
the control's $\sigma^x$:
\begin{align}
   \label{eq:Decoh_int_4}
   \rho'_{ \Sys \Control } & = \frac{1}{4} [
   ( V W_t  \ketbra{ \psi }{ \psi }  W_t V
      +  W_t  V  \ketbra{ \psi }{ \psi }  V  W_t
      \nonumber \\ & \qquad
      +  p V W_t  \ketbra{ \psi }{ \psi }  V  W_t
      +  p W_t V  \ketbra{ \psi }{ \psi }  W_t  V )
      \otimes \ketbra{+}{+}
      \nonumber \\ & \quad
   + (  V W_t  \ketbra{ \psi }{ \psi }  W_t V
         +  W_t  V  \ketbra{ \psi }{ \psi }  V  W_t
          \nonumber \\ & \qquad
         -  p V W_t  \ketbra{ \psi }{ \psi }  V  W_t
         -  p W_t  V  \ketbra{ \psi }{ \psi }  W_t  V  )
       \otimes  \ketbra{-}{-}
       \nonumber \\ & \qquad \;
   +  ( \text{cross-terms} ) ]  \, .
\end{align}

The control's $\sigma^x$ has the expectation value
\begin{align}
   \label{eq:Decoh_int_5}
   \expval{ X }_\Control^{ ( W, V , p ) }
   =  p  \,  \Re ( F_t )  \, .
\end{align}
The expectation value is proportional to the signal.
The ``not depolarized'' probability $p$
reduces the signal.

Consider repeating the protocol with $V = W = \id$.
The expectation value becomes
\begin{align}
   \label{eq:Decoh_int_6}
   \expval{ X }_\Control^{ ( \id , \id , p ) }
   =  p  \, .
\end{align}
Renormalizing the right-hand side of Eq.~\eqref{eq:Decoh_int_5} with
the right-hand side of Eq.~\eqref{eq:Decoh_int_6} yields
the OTOC's real part:
\begin{align}
   \label{eq:Decoh_int_7}  \boxed{
   \Re ( F_t )  =  \frac{
   \expval{ X }_\Control^{ ( W, V , p ) }  }{
   \expval{ X }_\Control^{ ( \id , \id , p ) } }
   }  \, .
\end{align}
The OTOC can be inferred perfectly, without approximation.
Furthermore, the not-depolarized probability $p$ can be inferred
in the absence of channel tomography,
which costs substantial time
and classical computation.

\subsection{Numerical simulations of decoherence}
\label{section:Decoher_Numerics}

To explore the physics of environmental decoherence numerically, we adopt the following simple model. We consider two equal-length chains of the power-law quantum Ising model, a system chain $\cal{S}$ and an environment chain $\cal{E}$. The Hamiltonian is
\begin{align}
H_{\cal{S}\cal{E}}
= H_{\cal{S}} + H_{\cal{E}}
+ J_\Coupling  \sum_{i=1}^{n_{\cal{S}}} \sigma^z_{i} \sigma^z_{i + n_{\cal{S}}}  \, ,
\end{align}
wherein $H_{\cal{S}}$ and $H_{\cal{E}}$ are power-law-quantum-Ising Hamiltonians, the system consists of qubits $\{1,...,n_{\cal{S}}\}$, and the environment consists of qubits $\{n_{\cal{S}}+1,...,2 n_{\cal{S}}\}$.
Each system qubit $i$ couples to
the corresponding environmental qubit $i$.

In the time-reversal procedure, the forward Hamiltonian is
\begin{align}
H_1 = H_{\cal{S}\cal{E}}
= H_{\cal{S}} + H_{\cal{E}}
+ J_\Coupling  \sum_{i=1}^{n_{\cal{S}}} \sigma^z_{i} \sigma^z_{i + n_{\cal{S}}}  \, ,
\end{align}
and the backward Hamiltonian is
\begin{align}
H_2 = H_{\cal{S}\cal{E}}
= H_{\cal{S}} - H_{\cal{E}}
- J_\Coupling \sum_{i=1}^{n_{\cal{S}}}  \sigma^z_{i} \sigma^z_{i + n_{\cal{S}}}  \, .
\end{align}
Only the system Hamiltonian is reversed.

\begin{figure}[hbt]
\centering
\includegraphics[width=.48\textwidth, clip=true]{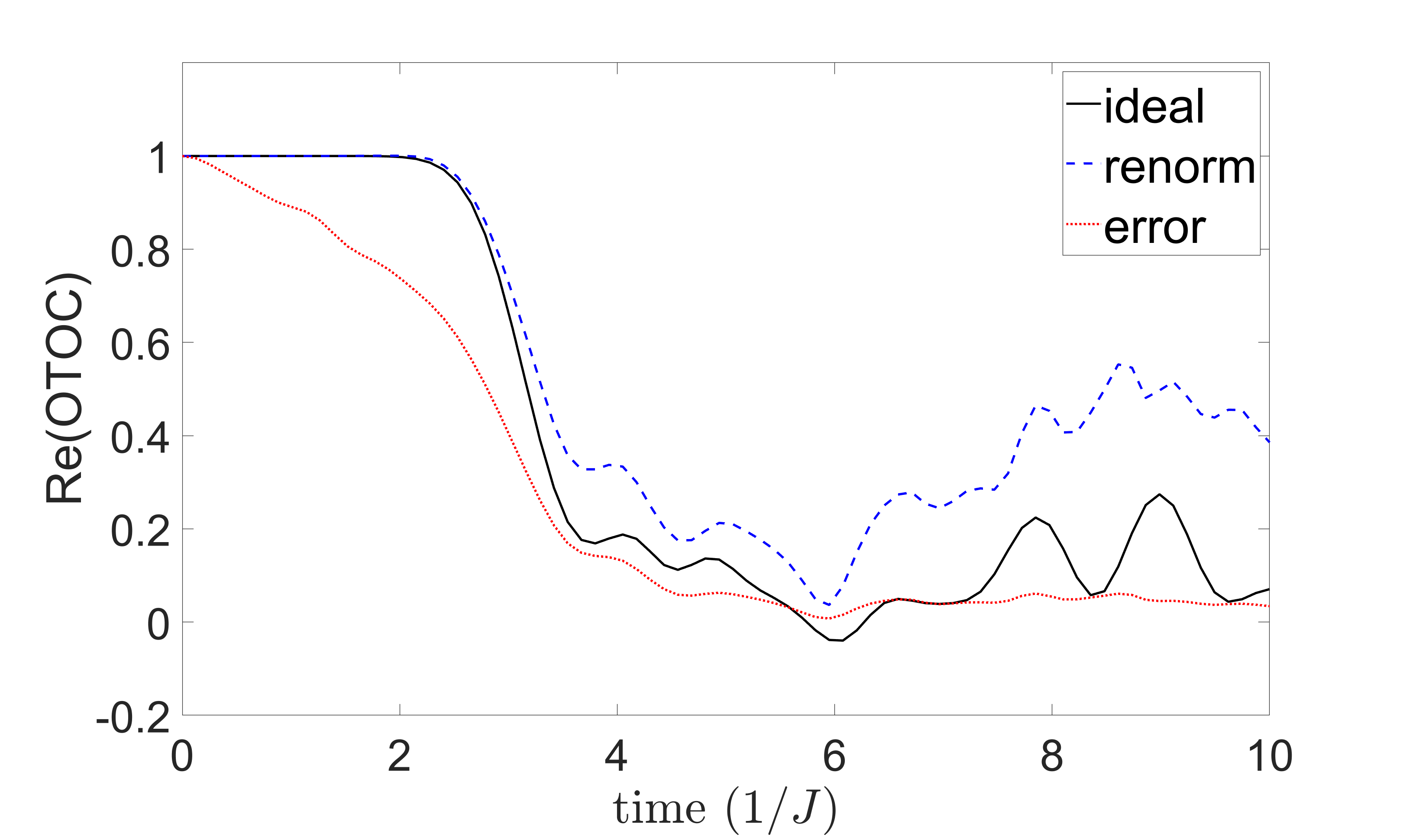}
\caption{\caphead{Open-system results:}
Power-law quantum Ising model with $n_{\cal{S}}=7$ spins (the system) coupled to
another power-law quantum Ising model with $n_{\cal{E}}=7$ spins (the environment),
via $\sigma^z \sigma^z$ couplings of strength $J_\Coupling = .2$.
The time-reversal procedure is defined by
a full reversal of the system Hamiltonian
without any change to
the environmental Hamiltonian or the coupling Hamiltonian.}
\label{fig:nS7Jc0p2_env}
\end{figure}

\begin{figure}[hbt]
\centering
\includegraphics[width=.48\textwidth, clip=true]{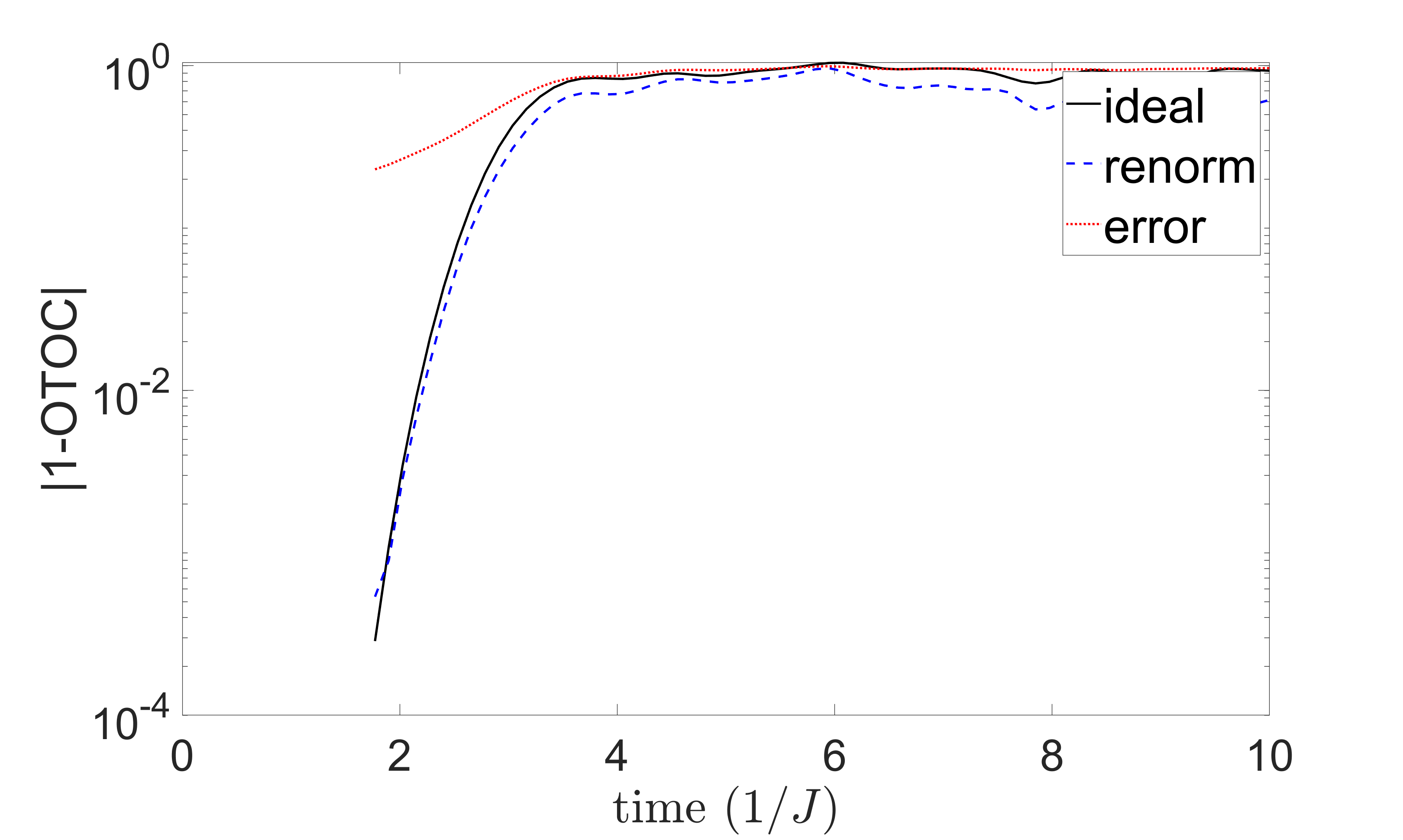}
\caption{\caphead{Open-system results:}
Same data as in Figure~\ref{fig:nS7Jc0p2_env}, on a semilogarithmic plot. }
\label{fig:nS7Jc0p2_env_log}
\end{figure}

Figures~\ref{fig:nS7Jc0p2_env} and \ref{fig:nS7Jc0p2_env_log} show the results of our interferometric renormalization scheme applied to this situation when $J_\Coupling=.2$. There is now significant deviation at early times on the semilogarithmic plot.
But, given how crude this time-reversal procedure is and how strong the coupling is, the agreement remains reasonably good. The early-time growth rate, as extracted from the renormalized data, is still much closer to the ideal result than the imperfect data is.

\begin{figure}[hbt]
\centering
\includegraphics[width=.48\textwidth, clip=true]{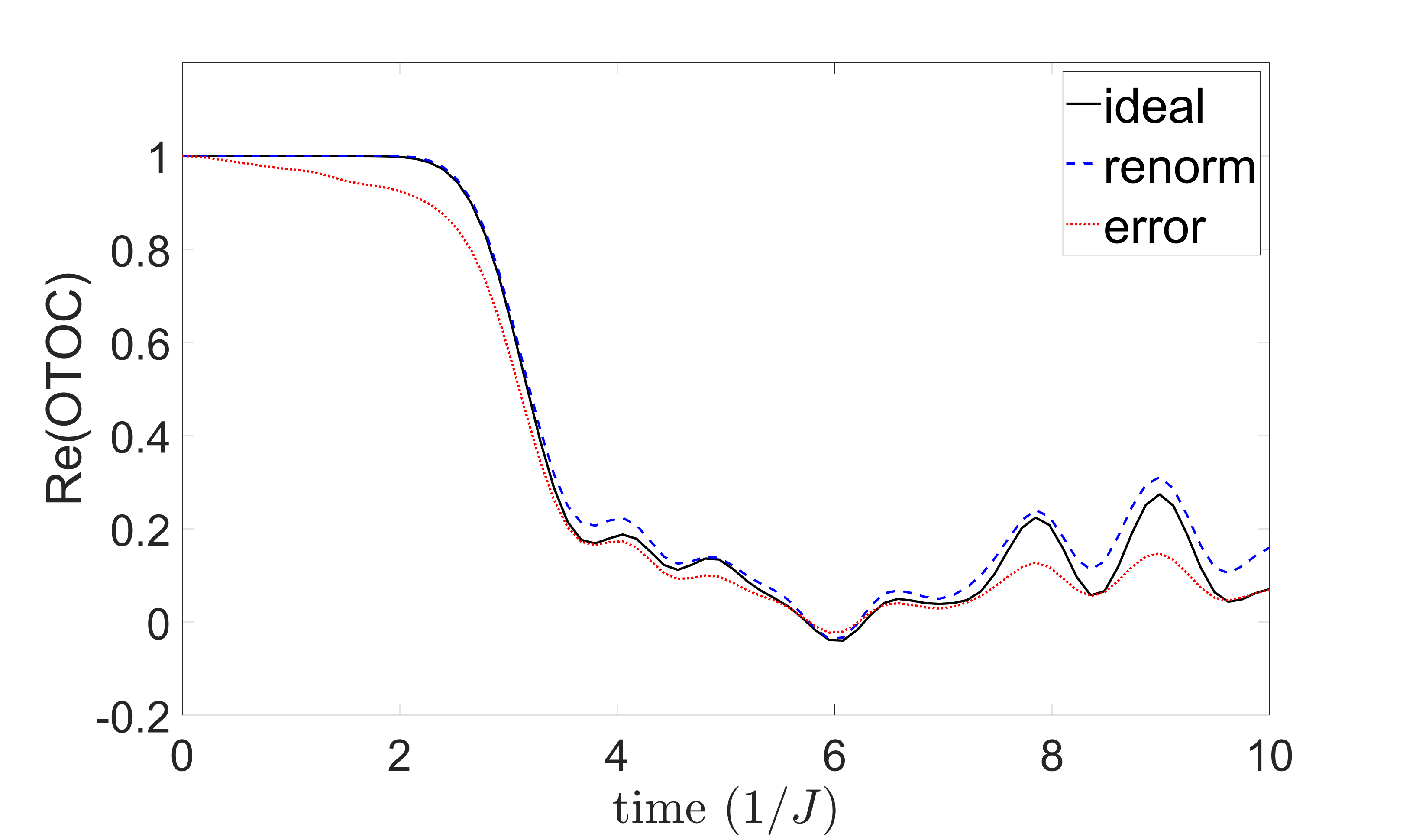}
\caption{\caphead{Open-system results:}
Power-law quantum Ising model with $n_{\cal{S}}=7$ spins (the system) coupled to
another power-law quantum Ising model with $n_{\cal{E}}=7$ spins (the environment)
via $\sigma^z \sigma^z$ couplings of strength $J_\Coupling = .1$.
The time-reversal procedure is defined by
a full reversal of the system Hamiltonian
without any change to
the environmental Hamiltonian or the coupling Hamiltonian.}
\label{fig:nS7Jc0p1_env}
\end{figure}

\begin{figure}[hbt]
\centering
\includegraphics[width=.48\textwidth, clip=true]{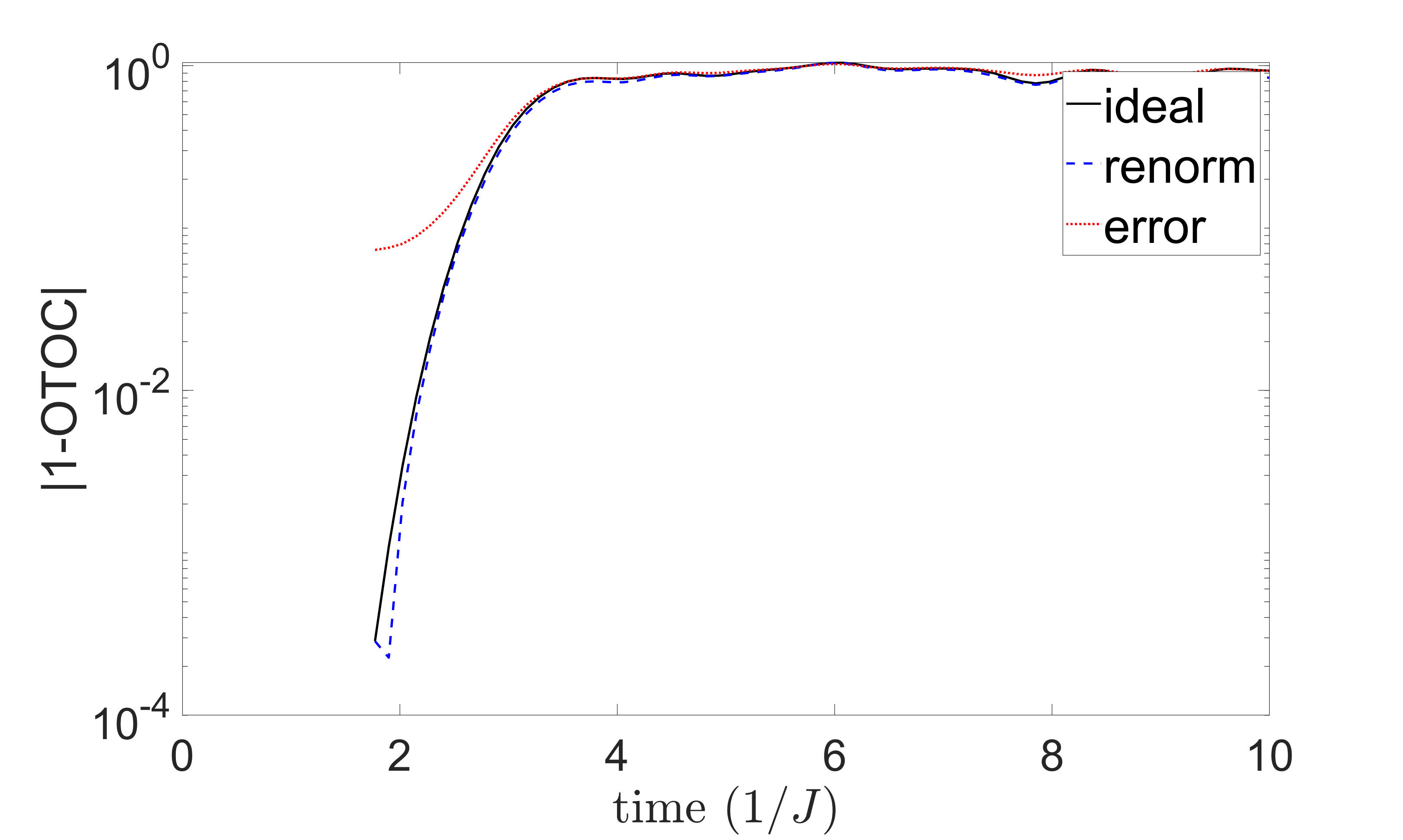}
\caption{\caphead{Open-system results:}
Same data as in Figure~\ref{fig:nS7Jc0p1_env}, on a semilogarithmic plot. }
\label{fig:nS7Jc0p1_env_log}
\end{figure}

Figures~\ref{fig:nS7Jc0p1_env} and \ref{fig:nS7Jc0p1_env_log} show the same scheme, with a reduced $J_\Coupling=.1$. Now, not only is the imperfect data relatively close to the ideal result,
but the renormalized data also cleaves very closely to the ideal result even well after the scrambling time for the small sizes considered here. So, while these models differ substantially from the simple depolarization channel in Sec.~\ref{section:Decoher_Analytics}, we find a similar conclusion about the renormalization scheme's efficacy in mitigating environmental decoherence.

\section{Holographic model}
\label{section:Holographic}

Let us show that the conclusions above
are not accidents of small system size,
of the models considered, or of infinite temperature.
We perform an analytical calculation in
a strongly chaotic system, using the holographic
anti-de-Sitter-space/conformal-field-theory (AdS/CFT) duality.
The renormalization formula holds for  simple timing errors,
even at finite temperatures,
up to the scrambling time.
The timing error is the simplest imperfection
that can studied holographically:
The forward and backward time evolutions last for
different lengths of time.
In the language above, $H_1$ are $H_2$ proportional,
but not generally equal, to $H$.

As stated, the goal is to show that the renormalization formula works in a highly nontrivial setting far beyond the system sizes explored in the numerical simulations. However, the calculation should not be viewed as a useless toy model:
Engineering a controlled quantum many-body system
that would exhibit a version of holographic duality
is a realistic experimental goal
(e.g.,~\cite{Garttner_16_Measuring,Swingle_17_Seeing,Bernien_17_Probing,Chen_18_Quantum}).
Such a system would allow experimental access to black-hole scrambling.
Hence it is sensible to assess
the robustness of scrambling measurements in
highly chaotic systems dual to gravity.

Let the forward-evolution time be $t_1 = t + \delta_1$,
and let the reverse time be $t_2 = t + \delta_2$.
If $\rho$ is a thermal state $e^{ - \beta H } / Z$,
the imperfect OTOC is
\begin{equation}
   \tilde{F}_t = \text{Tr}  (  W^\dagger_t V^\dagger_{-\delta_2} W_t V_{-\delta_1} \rho ),
\end{equation}
wherein, again, $O_t := e^{i H t} O e^{-i H t}$ is
a Heisenberg-picture operator.

Two simplifications prove convenient
in the holographic calculation:
First, $V$ is assumed to be Hermitian.
Second, we deform the OTOC to a \emph{thermally regulated OTOC}.
Thermal regulation does not change the essential physics of scrambling in this model. We consider a thermally regulated version of $\tilde{F}_t$ of the form
\begin{equation}
   \label{eq:Hol_1}
   \tilde{F}_t^{\text{reg}}
   = \text{Tr}   (  W^\dagger_t V_{-\delta_2} W_t  \sqrt{\rho}
   V_{-\delta_1} \sqrt{\rho} )  \, .
\end{equation}
Other thermal regulations are possible.
This choice is convenient because it captures the physics of scrambling and maps cleanly to a geometric problem.\footnote{
Consider a general thermal correlation function
$\langle A(t_1) B(t_2) C(t_3) D(t_4)\rangle$.
What we call ``thermal regulation'' amounts to
shifting some of the time arguments by imaginary terms.
The imaginary-time evolution operator
is proportional to a power of $e^{ - \beta H } / Z$.
This analytic continuation therefore amounts to
breaking $\rho$ into pieces
and distributing them amongst $A$, $B$, $C$, and $D$.
See, for example,~\cite{Maldacena_15_Bound}.}

$\tilde{F}_t^{\text{reg}}$
is related to the expectation value of the tensor product of $V$ with its transpose,
$V_{-\delta_2} \otimes \left(V_{-\delta_1}\right)^T$,
in a doubled system.
By ``doubled system,'' we mean
two copies of the system of interest.
The relevant whole-system state results from
having perturbed the thermofield double with $W$.
The \emph{thermofield double}
\begin{equation}
   |\text{TFD}\rangle = \sum_i
   \sqrt{\frac{e^{-\beta E_i}}{Z}} |E_i \rangle \otimes |E_i \rangle
\end{equation}
purifies the thermal $\rho$.
The perturbed thermofield double state is
\begin{equation}
|W\rangle = \left( W_t \otimes \id \right) |\text{TFD} \rangle  \, .
\end{equation}
Hence
\begin{equation}
   \tilde{F}_t^{\text{reg}}
   = \langle W | V_{-\delta_2} \otimes \left(V_{-\delta_1}\right)^T |W \rangle.
\end{equation}
We define the transpose using the energy basis, such that
$\left(O_t \right)^T= \left( O^T\right)_{-t}$ and
\begin{equation}
   \tilde{F}_t^{\text{reg}}
   = \langle W | V_{-\delta_2} \otimes \left(V^T\right)_{\delta_1} |W \rangle.
\end{equation}
This expectation value is related,
via the AdS/CFT duality, to
a correlation function between
the two sides of
an eternal black hole
perturbed by a shock wave caused by $W$.

Assume that the shock wave does not add much energy to the system.
The bulk geometry is described by
a mass-$M$ black hole
perturbed, on the horizon, by a shock wave with a null shift $\alpha$.
$M$ is determined by the thermal-state temperature $1 / \beta$
(we set Boltzmann's constant to $\kB = 1$).
The details of this geometry are recorded in~\cite{Shenker_Stanford_14_BHs_and_butterfly}.
Let $t=-t_w$ denote the long-ago time at which $W$ perturbed the system.\footnote{
$t$ is often assumed to be positive.
But the same physics results for negative $t$
in the thermal state, if $W$ and $V$ are exchanged.
Since this model's scrambling physics
does not depend strongly on $W$ and $V$,
we are free to choose the most convenient sign for $t$.}
Let $\delta E$ denote the energy added to the system by $W$.
In a convenient Kruskal coordinate system,
the perturbation shifts
the coordinates in the left-hand geometry
relative the right-hand coordinates by an amount
$\alpha = \frac{\delta E}{4M}e^{2\pi t_w/\beta}$.

$\tilde{F}_{t}^{\text{reg}}$ will be analyzed in a geodesic approximation.
Consider the two boundary points at which the $V$ operators are inserted.
The renormalized geodesic distance between these points is
\begin{equation}
   \frac{d}{\ell}\bigg|_{\text{ren}}
   = 2 \log \left(  \cosh  \left(  \frac{\pi ( t_\Left  -  t_\Right )}{\beta}   \right)
   + \frac{\alpha}{2} e^{-\pi(  t_\Left + t_\Right)/\beta} \right),
\end{equation}
wherein $\ell$ denotes the AdS radius, Planck's constant $\hbar = 1$, and
$t_\Left$ and $t_\Right$ denote the times at which
the $V$'s operate on the left and right boundaries.
In our case, $t_\Left = \delta_2$, and $t_\Right  =  \delta_1$.
``Renormalized'' refers, here, to the removal of field-theory divergences,
not to the renormalization formula~\eqref{eq:FInt_Conjecture}.
In fact, the field-theory renormalizations cancel from
the renormalization formula's numerator and denominator.

Let $V$ be a primary field with dimension $\Delta$
(and bulk mass $\sim \Delta/\ell$).
The geodesic approximation to the correlator is
\begin{equation}
   \tilde{F}_t^{\text{reg}}
   \sim \left(\frac{1}{ \cosh  \left( \frac{\pi (\delta_2-\delta_1)}{\beta}  \right)
   + \frac{\alpha}{2} e^{-\pi(\delta_2+\delta_1)/\beta}} \right)^{2\Delta}.
\end{equation}
Let us expand in small $\alpha$, as is reasonable until just before
the scrambling time, $t_*$:
\begin{eqnarray}
   \tilde{F}_t^{\text{reg}}
   \sim & \left(  \frac{1}{ \cosh \left(  \frac{\pi (\delta_2-\delta_1)}{\beta}  \right) } \right)^{2\Delta}  \nonumber \\
   &  \times \left(1 - \Delta \alpha
   \frac{e^{-\pi (\delta_1 + \delta_2)/\beta}}{
           \cosh \left(  \frac{\pi (\delta_2-\delta_1)}{\beta}  \right) }
   + \cdots \right).
\end{eqnarray}

Typically, many experimental shots are required to build up enough statistics
to estimate the value of $F_t$.
This process will be complicated if
the values of $\delta_1$ and $\delta_2$ vary from shot to shot.
The sensible thing to do is to
(i) average over shots,
to estimate the values of $\tilde{F}_t(W,V)$ and $\tilde{F}_t( \id ,  V)$ separately,
and then (ii) take the ratio to estimate $F_t$ via the renormalization formula.
Would such a procedure yield nearly the correct value of $F_t$?

\subsection{Simple error distribution}

Let $\delta_i = \pm \epsilon t_w$ with probability $1/2$ for $i = 1,2$:
In every shot, the system 
has a probability $1/2$ of being over-evolved for
a fraction $\epsilon$ of the total time
and a probability $1/2$ of being under-evolved analogously.
To reduce notation, we relabel the renormalization-formula numerator as
$A_1 = \tilde{F}_t^{\text{reg}}(W,V)$
and the denominator as
$A_2=\tilde{F}_t^{\text{reg}}(\id, V)$.
The shot-average of $A_2$ is
\begin{equation}
   \overline{A_2} = \frac{1}{2}
   + \frac{1}{2}
   \frac{1}{\left[\cosh  \left(   \frac{2\pi \epsilon t_w}{\beta}  \right)  \right]^{2\Delta}}.
\end{equation}
Similarly, the shot-average of $A_1$, to leading order in $\alpha$, is
\begin{align}
   \overline{A_1} = \overline{A_2}
   - \frac{\Delta \alpha }{2}
   \left[  \frac{1}{\left[\cosh  \left(  \frac{2\pi \epsilon t_w}{\beta}  \right)  \right]^{2\Delta+1}}
   + \cosh \left(  \frac{2\pi \epsilon t_w}{\beta}  \right)  \right].
\end{align}
We can check the limit as $\epsilon t_w \rightarrow 0$:
$\overline{A_2} \rightarrow 1$, and
$\overline{A_1} \rightarrow 1-\Delta \alpha$, which are the ideal values.
The renormalized value for general $\epsilon t_w$ but small $\alpha$ is
\begin{equation}
   1 - \Delta \alpha
   \frac{\left[   \cosh  \left(  \frac{2\pi \epsilon t_w}{\beta}  \right) \right]^{2\Delta+2} + 1  }{\left[ \cosh \left(  \frac{2\pi \epsilon t_w}{\beta}  \right) \right]^{2\Delta+1}
   + \cosh  \left(  \frac{2\pi \epsilon t_w}{\beta}  \right)  } \, .
\end{equation}

Suppose that the timing error is severe: $\epsilon t_w \gg \beta$.
The measured correlators limit as
$\overline{A_2}\rightarrow 1/2$ and
$\overline{A_1} \rightarrow 1/2-\Delta \alpha e^{2\pi \epsilon t_w/\beta}/4 + \ldots$
The renormalization formula becomes
\begin{equation}
   \frac{ \:  \;  \overline{A_1}  \;  \: }{\overline{A_2}}
   \to 1 - \Delta \alpha \frac{e^{2\pi \epsilon t_w/\beta}}{2}
   +  \ldots
\end{equation}
Recall that (i) the ideal value is $F = 1 - \Delta \alpha + \cdots$ and
(ii) $\alpha = \frac{\delta E}{4M}e^{2\pi t_w/\beta}$.
Substituing shot-averaged quantities
into the renormalization formula
therefore gives exponential growth.
The exponent differs from the ideal value
by no more than a factor of $\epsilon$.

\begin{figure}
\begin{center}
\includegraphics[width=.49\textwidth]{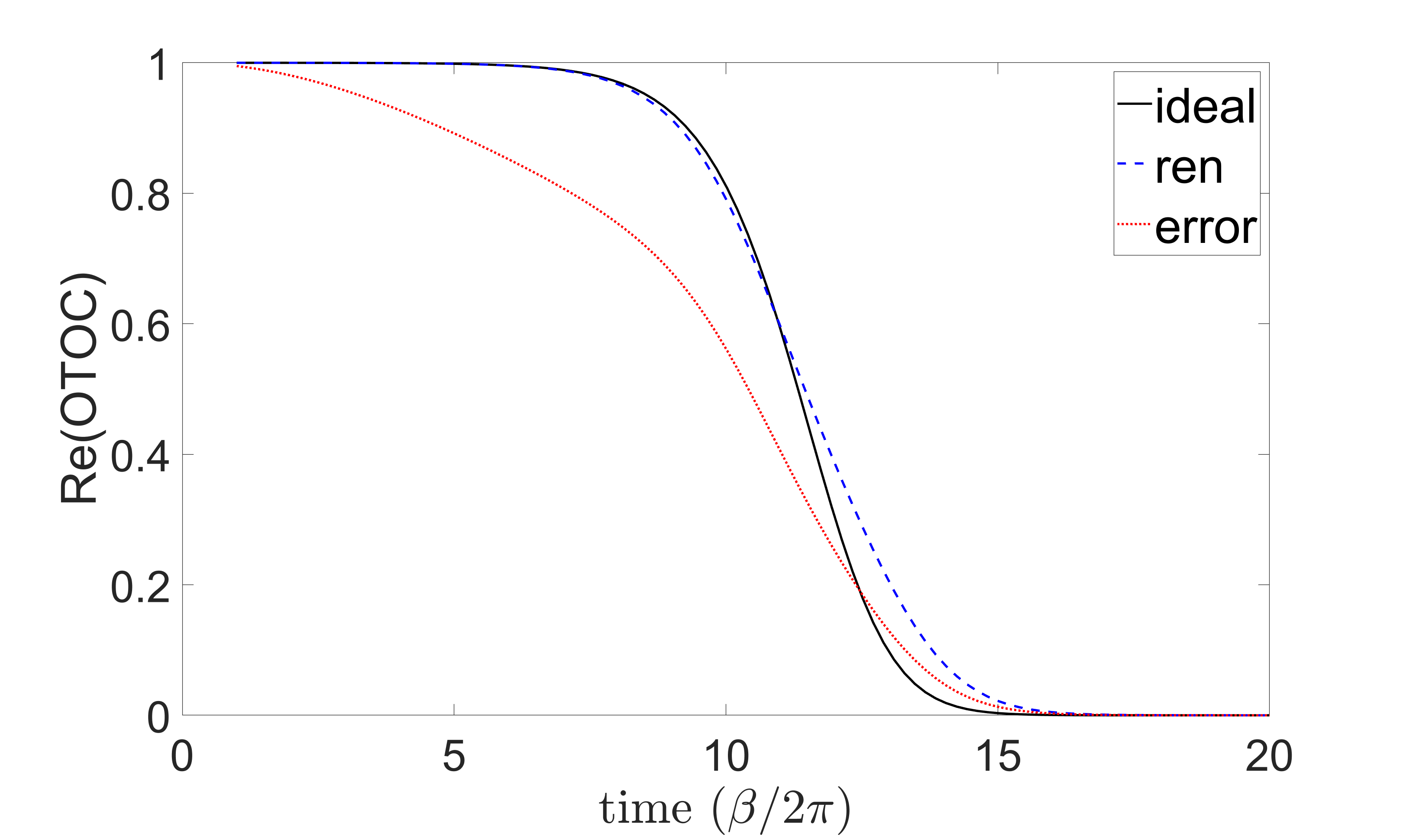}
\caption{\caphead{Renormalization scheme in
strongly holographic model:}
The shot-averaged regulated out-of-time-ordered correlator $\tilde{F}^\reg_t$
is plotted against $t$, measured in units of $(\beta/2\pi)$.
The null shift   $\alpha = G e^{2\pi t/\beta}$.
The ratio $G = \frac{ \delta E }{ 4 M }$
is set to $G=10^{-5}$:
The perturbation is tiny.
The timing error is 10$\%$: $\epsilon=0.1$.
Black represents the ideal $\tilde{F}^\reg_t$;
blue dashed, the renormalized value;
and red dotted, the unrenormalized imperfect value.}
\label{fig:ads3_fig1}
\end{center}
\end{figure}

\begin{figure}
\begin{center}
\includegraphics[width=.49\textwidth]{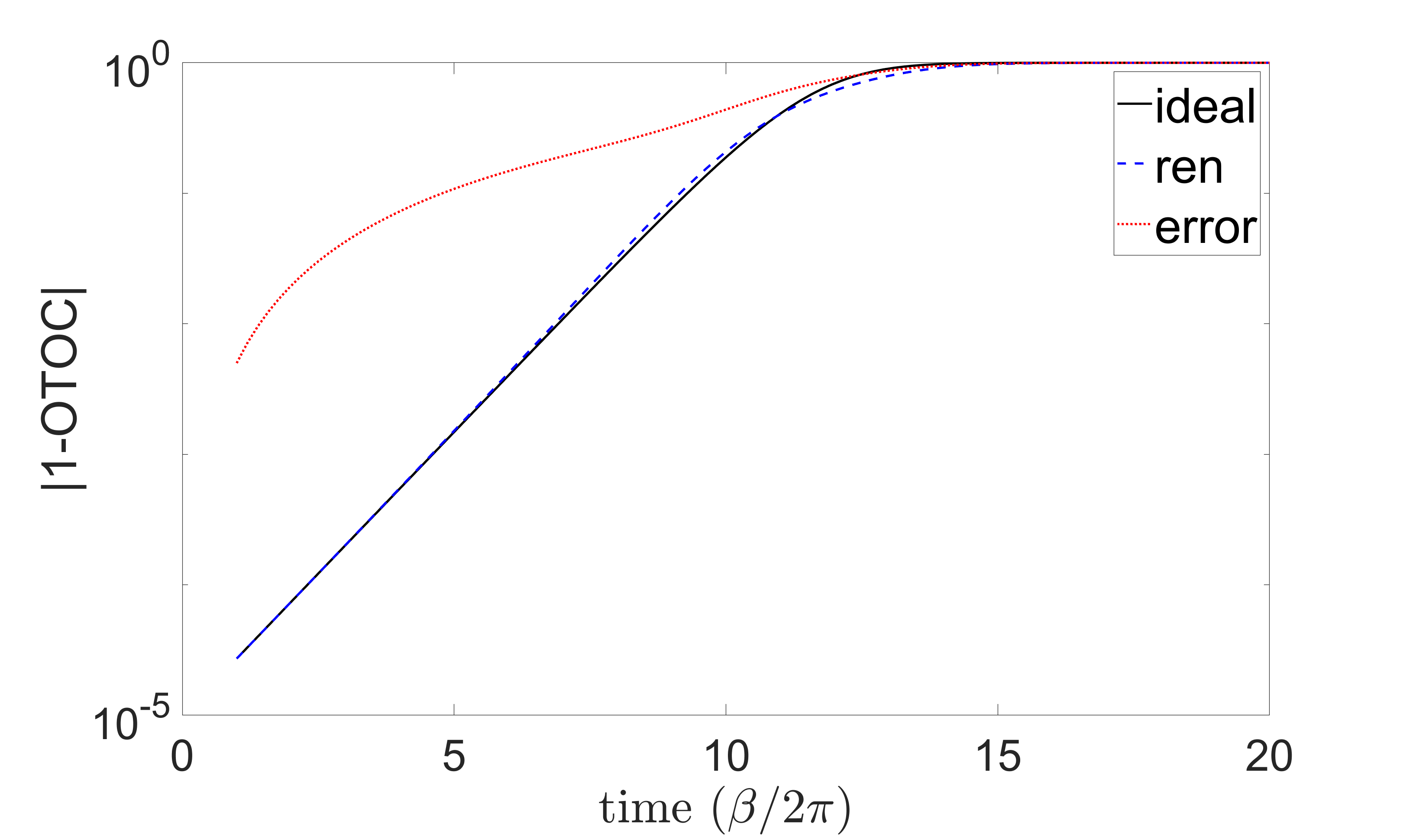}
\caption{\caphead{Renormalization scheme in
strongly holographic model:}
Same parameters and data as in Figure~\ref{fig:ads3_fig1}, on a logarithmic scale.} \label{fig:ads3_fig1_log}
\end{center}
\end{figure}

We can also study the renormalization scheme away from small $\alpha$.
The general results are
\begin{equation}
   \overline{A_2} = \frac{1}{2}
   + \frac{1}{2}
   \frac{1}{\left[\cosh  \left(  \frac{2\pi \epsilon t_w}{\beta}  \right)  \right]^{2\Delta}}
\end{equation}
and
\begin{align}
   \overline{A_1}
   = & \frac{1}{4} \left(\frac{1}{1 + \frac{\alpha}{2} e^{-2\pi \epsilon t_w/\beta}}\right)^{2\Delta} 
   + \frac{1}{4} \left(\frac{1}{1 + \frac{\alpha}{2} e^{+2\pi\epsilon t_w/\beta}}\right)^{2\Delta} \nonumber \\
   & + \frac{1}{2}  \left(\frac{1}{\cosh  \left( \frac{2\pi \epsilon t_w}{\beta}  \right)
   + \frac{\alpha}{2} }\right)^{2\Delta}.
\end{align}
These results are illustrated Figures~\ref{fig:ads3_fig1} and \ref{fig:ads3_fig1_log}.
The scheme's quality is excellent even for a 10$\%$ timing error
($\epsilon = 0.1$).
More precisely, the correct exponential growth is encoded in
$\alpha  \sim  e^{2\pi t/\beta}$.
The renormalization formula predicts
an exponential growth of $e^{2\pi(1+\epsilon)t/\beta}$.
Hence even in this strongly chaotic model of many degrees of freedom\footnote{
Let us reparameterize the null shift as
$\alpha  \equiv  G e^{ 2 \pi t / \beta }$,
wherein $G  :=  \frac{ \delta E }{ 4 M }$.
The entropy $S$ scales as $1/G$, wherein
$G$ plays the role of Newton's constant.
(Assume that the energy perturbation obeys $\delta E \sim 1/\beta$.
Since $ M \sim  S/\beta$,
$\frac{\delta E}{M} \sim \frac{1}{S}$.
Hence $G  \sim  \frac{ \delta E }{ M }  \sim  \frac{1}{S}$.
Similarly, the entropy $S_{\rm BH}$ of a general black hole
varies inversely with Newton's constant:
$S_{\rm BH} \propto \frac{1}{ G_{\rm N} }$;
hence our use of the notation $G$.)}
at finite temperature,
the renormalization scheme estimates
the correct exponent to relative error of order $\epsilon$.

\section{Conclusions}
\label{section:Conclusions}

We have shown, with analytical arguments and numerical simulations,
that scrambling measurements are remarkably resilient with respect to
imperfections in the experimental protocol.
Our physical interpretation of the results is that
the physics of scrambling can be cleanly separated from
the decay of fidelity due to imperfections, up to the scrambling time.
We exhibited this resilience for a chaotic local spin chain of up to 
$n=18$ sites and for
a strongly chaotic holographic model with many degrees of freedom.
We have checked that our conclusions apply also many other models.
Examples include integrable models and
nonlocal models (e.g., the Sachdev-Ye-Kitaev (SYK) model~\cite{Sachdev_93_Gapless,Kitaev_15_Simple,Polchinski_16_Spectrum,Maldacena_16_Comments}).
We focused on states near the energy spectrum's center.
But the renormalization scheme applies to other states, e.g., the ground state.
Thus, the resilience of scrambling measurements shown here
is quite general.

In the numerical analysis, we considered mostly modest system sizes.
The choice facilitates the study of many models and set-ups with
a reasonable amount of computer time.
We studied a few larger system sizes, however---up to $n=20$ spins.
We found, at most, a modest degradation in
the renormalization scheme's effectiveness
until the scrambling time.
Precisely how the renormalization scheme's effectiveness
scales with $n$ remains an open question. The holographic analysis, which applies to a system with many degrees of freedom, gives evidence of a favorable scaling with system size.
Experiments should be able to create headway.

Perhaps our results' most important consequences are for experiments. Our renormalization schemes are simple and general and should greatly enhance early experiments' abilities to probe the physics of scrambling. For example, imperfections in the time-reversal scheme appear readily addressable with our methods. To that end, it would be very interesting to study in detail our renormalization scheme, with realistic assumptions, in the context of various near term experimental platforms.

Along these lines, one unrealistic assumption made in
the numerical analysis was that
the imperfections were the identical in all experimental runs.
We lift this assumption in Appendix~\ref{section:ShotToShot}:
The renormalization formula, phrased in terms of shot-averaged quantities,
remains valid despite shot-to-shot variations in the imperfections.

Our results also enable the use of new approximate time-reversal schemes.
For example, consider reversing only the fields and
the odd-index-neighbor couplings in the power-law quantum Ising model.
This scheme may seem artificial.
But consider an experiment in which local fields are easy to control
but the interactions are fixed.
Local unitary transformations and field reversal
can effect such a partial time reversal.
Such a reversal, combined with our renormalization scheme,
gives excellent agreement with the ideal-time-reversal results.

Testing the scheme in larger experimental systems
would help illuminate our renormalization scheme's physics.
Indeed, the quantum physics of near-term noisy quantum devices
presents an exciting frontier today~\cite{Preskill_18_Quantum}.
Our results suggest that scrambling
might be amenable to study on noisy near-term machines.
Relatedly, a similar procedure of dividing by a Loschmidt echo
has been used in analysis of nuclear-magnetic-resonance experiments~\cite{Sanchez_16_NMRMQCReview}.

In our quest to better understand our resilience results' significance,
calculations in model systems will be valuable.
The numerics here form a black-box approach.
More insight may come from opening the box,
budding off from the holographic calculation in Sec.~\ref{section:Holographic} and
the decoherence models in Sec.~\ref{section:Decoher}.
Perhaps the physics of scrambling resilience can be related to known types of robustness, e.g., the robustness of renormalization-group fixed points.
It would be interesting to probe resilience in
many other recently studied models,
including noninteracting, weakly coupled, and semiclassical systems~\cite{Stanford_15_WeakCouplingChaos,Patel_16_ChaosCritFS,Patel_17_DisorderMetalChaos,Aleiner_16_Microscopic,Chowdhury_17_ONChaos,Lin_18_Out},
many-body-localized states~\cite{Huang_16_MBL_OTOC,Fan_16_MBL_OTOC,He_16_MBL_OTOC,Chen_16_MBL_OTOC,Swingle_16_MBL_OTOC},
the SYK model~\cite{Sachdev_93_Gapless,Kitaev_15_Simple,Polchinski_16_Spectrum,Maldacena_16_Comments}, open systems~\cite{Syzranov_17_Out},
local random-circuit models~\cite{Nahum2017a,VonKeyserlingk2017,Khemani2017,Rakovszky2017}, other special solvable models~\cite{Tsuji_17_Exact}, and much else.

Finally, an extension of the renormalization scheme
to the out-of-time-ordered-correlator (OTOC) quasiprobability
$\SumKD{\rho}$ merits further study.
Two approaches suggest themselves:
(i) The analytical argument
of Sec.~\ref{section:Weak_PseudoDerivn} might be modified:
Projectors $\ProjW{w_\ell}$ and $\ProjV{v_\ell}$
might replace the unitaries $W$ and $V$.
Yet $\ProjV{v_\ell}$ lacks the unitary property $V^\dag V = \id$.
Perhaps this lack can be circumvented.
(ii) Suppose that the eigenvalues of $W$
and the eigenvalues of $V$ equal $\pm 1$.
(Suppose, for example, that $W$ and $V$ are Paulis.)
$\SumKD{\rho}$ equals a combination of
$F_t$ and simpler correlators~\cite[Sec.~II D]{NYH_17_Quasi}.
$F_t$ can be renormalized, we have shown.
Each simpler correlator needs no renormalization,
or appears to be renormalizable generally,
or appears to be renormalizable under certain conditions on $\rho$
(e.g., if $\rho = \id / \Dim$).
Renormalizing every term, then assembling the terms,
is expected to yield a renormalized OTOC quasiprobability.

%
%
\begin{acknowledgments}
NYH is grateful for funding from the Institute for Quantum Information and Matter, an NSF Physics Frontiers Center (NSF Grant PHY-1125565) with support of the Gordon and Betty Moore Foundation (GBMF-2644);
for partial support from the Walter Burke Institute for Theoretical Physics at Caltech;
for a Graduate Fellowship from the Kavli Institute for Theoretical Physics;
for a Barbara Groce Graduate Fellowship;
and to Justin Dressel for weak-measurement discussions.
BGS is supported by the Simons Foundation, through the ``It From Qubit Collaboration,''
and by the National Science Foundation, under Grant
No. NSF PHY-1125915, and acknowledges useful discussions with Monika Schleier-Smith and Norm Yao.
\end{acknowledgments}

\begin{appendices}


\renewcommand{\thesection}{\Alph{section}}
\renewcommand{\thesubsection}{\Alph{section} \arabic{subsection}}
\renewcommand{\thesubsection}{\Alph{section} \arabic{subsection}}
\renewcommand{\thesubsubsection}{\Alph{section} \arabic{subsection} \roman{subsubsection}}

\makeatletter\@addtoreset{equation}{section}
\def\theequation{\thesection\arabic{equation}}

\section{Further motivation for renormalization of the interferometer:
Infinite-temperature analysis}
\label{section:TInfty}

Consider inputting an infinite-temperature state,
$\rho  =  \id / 2^\Sites$,
into the imperfect interferometer:
\begin{align}
F^{\text{int}}_t = \frac{1}{2^n} \text{Tr}\left(U_1^\dagger W^\dagger U_2 V^\dagger U_2^\dagger W U_1 V \right).
\end{align}
Define $V_i := U^\dagger U_i V U_i^\dagger U$, such that
\begin{align}
F^{\text{int}}_t = \frac{1}{2^n} \text{Tr}\left(W^\dagger_t V_2^\dagger W_t  V_1 \right).
\end{align}
Consider inserting an identity operator $\id = V^\dagger V$
leftward of the $V_2^\dag$:
\begin{align}
   F^{\text{int}}_t = \frac{1}{2^n} \text{Tr}\left(
   W^\dagger_t V^\dagger
   \left[ V V_2^\dagger \right]  W_t V_1 \right).
\end{align}
Since $\left(  V V_2^\dagger  \right) W_t
   = W_t \left(  V V_2^\dagger  \right)
   + \left[  V V_2^\dagger ,   W_t  \right]$,
\begin{align}
   F^{\text{int}}_t
   & = \frac{1}{2^n} \text{Tr}\left(
   W^\dagger_t V^\dagger W_t V
   \left[  V_2^\dagger V_1  \right] \right)
   \nonumber \\ & \quad
   + \frac{1}{2^n} \text{Tr}\left(W^\dagger_t V^\dagger
   \left[ V V_2^\dagger ,   W_t  \right]  V_1 \right) .
\end{align}

We can motivate the renormalization scheme by approximating the first term as
\begin{align}
& \frac{1}{2^n} \text{Tr}\left(
   W^\dagger_t V^\dagger W_t V
   \left[  V_2^\dagger V_1  \right] \right) \nonumber \\
   & \approx \frac{1}{2^n} \text{Tr}\left(
   W^\dagger_t V^\dagger W_t V \right)
   \frac{1}{2^n} \text{Tr}\left( V_2^\dagger V_1   \right),
\end{align}
and approximating the second term as
\begin{align}
\frac{1}{2^n} \text{Tr}\left(W^\dagger_t V^\dagger
   \left[ V V_2^\dagger ,   W_t  \right]  V_1 \right) \approx 0.
\end{align}
The first approximation is motivated by the fact that
it becomes exact as $W \rightarrow \id$ or if $[W_t, V] \approx 0$.
Hence the approximation is expected to be good
until roughly the scrambling time.

The second approximation is motivated by the fact that
matrix elements of commutators---objects
of the form $\frac{1}{2^n} \text{Tr}(A[B,C])$---are
generically small in chaotic, and in perturbed integrable, systems.
More precisely, consider early times at which,
by Trotter-expanding in the perturbation strength $\varepsilon$,
one can approximate $V_i \approx V + O(\varepsilon)$.
The second term should be smaller than the signal by
at least a factor of $\varepsilon$.

At later times, approximating $V_i \approx V$
is no longer possible.
By typical matrix elements of commutators
are expected to be small due to chaos---inherent
or arising from perturbed integrability.
One can object that $\frac{1}{2^n}\text{Tr}(W_t^\dagger V^\dagger [W_t,V])$ and $\frac{1}{2^n}\text{Tr}(V^\dagger W_t^\dagger [W_t,V])$ approach $\mp 1$, respectively, at late times in a chaotic system.
These examples appear to violate expectations.
This anomaly arises, however, because the operators inside and outside the commutator
are finely attuned to each other.
This tuning is absent from the second term above.

Even away from infinite temperature, aspects of the above discussion can be imitated. Consider feeding the perturbed interferometer a general pure state $|\psi \rangle$:
\begin{align}
   F^{\text{int}}_t
   = \langle \psi | U_1^\dagger W^\dagger U_2
   V^\dagger U_2^\dagger W U_1 V | \psi \rangle.
\end{align}
Let $|\tilde{\psi}\rangle := U^\dagger U_1 | \psi \rangle$, such that
\begin{align}
   F^{\text{int}}_t =\langle \tilde{\psi} | W_t^\dagger V_2^\dagger W_t V_1 | \tilde{\psi} \rangle.
\end{align}
Repeating the infinite-temperature analysis suggests that
\begin{align}
   & F^{\text{int}}_t \nonumber
    \approx \langle \tilde{\psi} | W_t^\dagger V^\dagger W_t V | \tilde{\psi} \rangle \langle \tilde{\psi} | V_2^\dagger V_1 | \tilde{\psi} \rangle.
\end{align}
The second term is
\begin{align}
\langle \tilde{\psi} | V_2^\dagger V_1 | \tilde{\psi} \rangle = F^{\text{int}}_t(\id,V),
\end{align}
the denominator in the renormalization scheme.

The first term has an appealing OTOC form,
but $\ket{ \tilde{\psi} }$ has replaced $\ket{ \psi }$.
How are the states' OTOCs related?
In a chaotic system,
any thermalized state's energy density
is expected to determine
the state's scrambling physics
in the thermodynamic limit.
Hence we must ask (i) is $|\tilde{\psi} \rangle$ a thermalized state
and (ii) how does the energy density of $|\tilde{\psi} \rangle$
differ from that of $| \psi \rangle$?

By late times---as the commutator-squared $| [ W(t) , V ] |^2$
grows appreciably---we expect $|\tilde{\psi}\rangle$ to be thermalized with respect to Hamiltonian $H$.
After all, the state has evolved under $H$ for
a long (negative) time.

Furthermore, we expect the state's average energy to be
$\langle \tilde{\psi} | H | \tilde{\psi} \rangle \approx \langle \psi | H_1 | \psi \rangle$.
To see why, think of $|\tilde{\psi}\rangle$ as arising from two evolutions.
$U_1$ governs the first evolution; and $U^\dagger$, the second.
As $U^\dag$ evolves the system,
the expectation value of $H$ is conserved:
\begin{align}
\langle \tilde{\psi} | H | \tilde{\psi} \rangle = \langle \psi | U_1^\dagger H U_1 | \psi \rangle.
\end{align}
The Hamiltonian decomposes as $H = H_1 + (H- H_1)$.
The $U_1$ evolution generically conserves
only the first term.
(Other conserved quantities can affect the analysis,
but we neglect this complication.)
Suppose that the $H_1$ evolution is chaotic.
(Even when $H$ is integrable, we expect
the typical perturbation not to be.)
The expectation value of $H-H_1$ will decay with time. Hence
\begin{align}
\langle \tilde{\psi} | H |\tilde{\psi} \rangle \approx \langle \psi | H_1 | \psi \rangle.
\end{align}

In the thermodynamic limit, the energy density should control the scrambling dynamics, e.g., by setting the effective system temperature.
Suppose that $H_1$ differs from $H$ by a systematic deviation of order $\varepsilon$.
The energy density of $|\tilde{\psi}\rangle$ should differ from
the energy density of $|\psi \rangle$ by
an amount of order $\varepsilon$.
This result constitutes the worst case.
Suppose now that, as in the numerical examples studied above, $H_1$ differs from $H$ by a random local deviation.
The total difference in energy is expected to be
proportional to $\sqrt{n}$, instead of to $n$.
The difference in energy density is of order $\varepsilon/\sqrt{n}$,
which vanishes in the thermodynamic limit.

This analysis suggests that, even away from infinite temperature,
the renormalization scheme reproduces
the scrambling physics of
a state whose energy density
differs from that of $|\psi \rangle$
by no more than $\varepsilon$.
Furthermore, if $H_1-H$ and $H_2 - H$ are sums of random terms,
the effective energy density is not expected to differ from the actual
in the thermodynamic limit.
In this case, the renormalization scheme could reproduce
the correct energy density's
ideal scrambling dynamics.

These arguments provide some theoretical motivation
for the renormalization scheme.
But the renormalized numerics' quality,
up to the scrambling time,
suggests to us that more remains to be discovered about
why the scheme works.

\section{Shot-to-shot imperfections}
\label{section:ShotToShot}

\begin{figure}[hbt]
\centering
\includegraphics[width=.48\textwidth, clip=true]{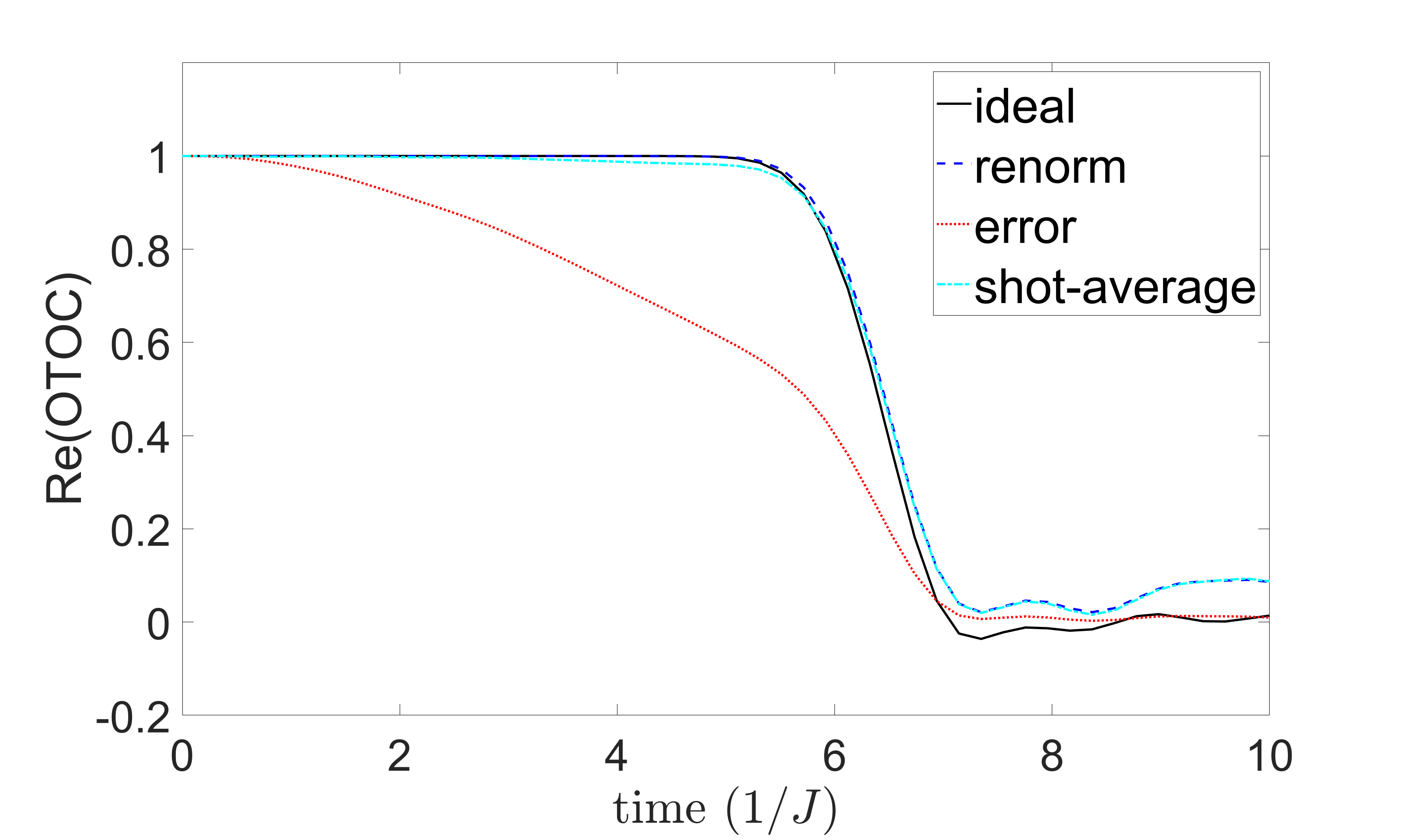}
\caption{\caphead{Shot-to-shot fluctuations:}
Floquet version of the power-law quantum Ising model.
The $\sigma^z$ terms were pulsed on
for a time interval $dt\approx .20$;
then the $\sigma^x$ terms were;
and so on, alternately.
The system consists of $n=12$ spins.
The imperfections fluctuate from shot to shot.
The shot-averaged quantities were computed from 100 samples.}
\label{fig:n12eps0p2_floq_manyshot}
\end{figure}

\begin{figure}[hbt]
\centering
\includegraphics[width=.48\textwidth, clip=true]{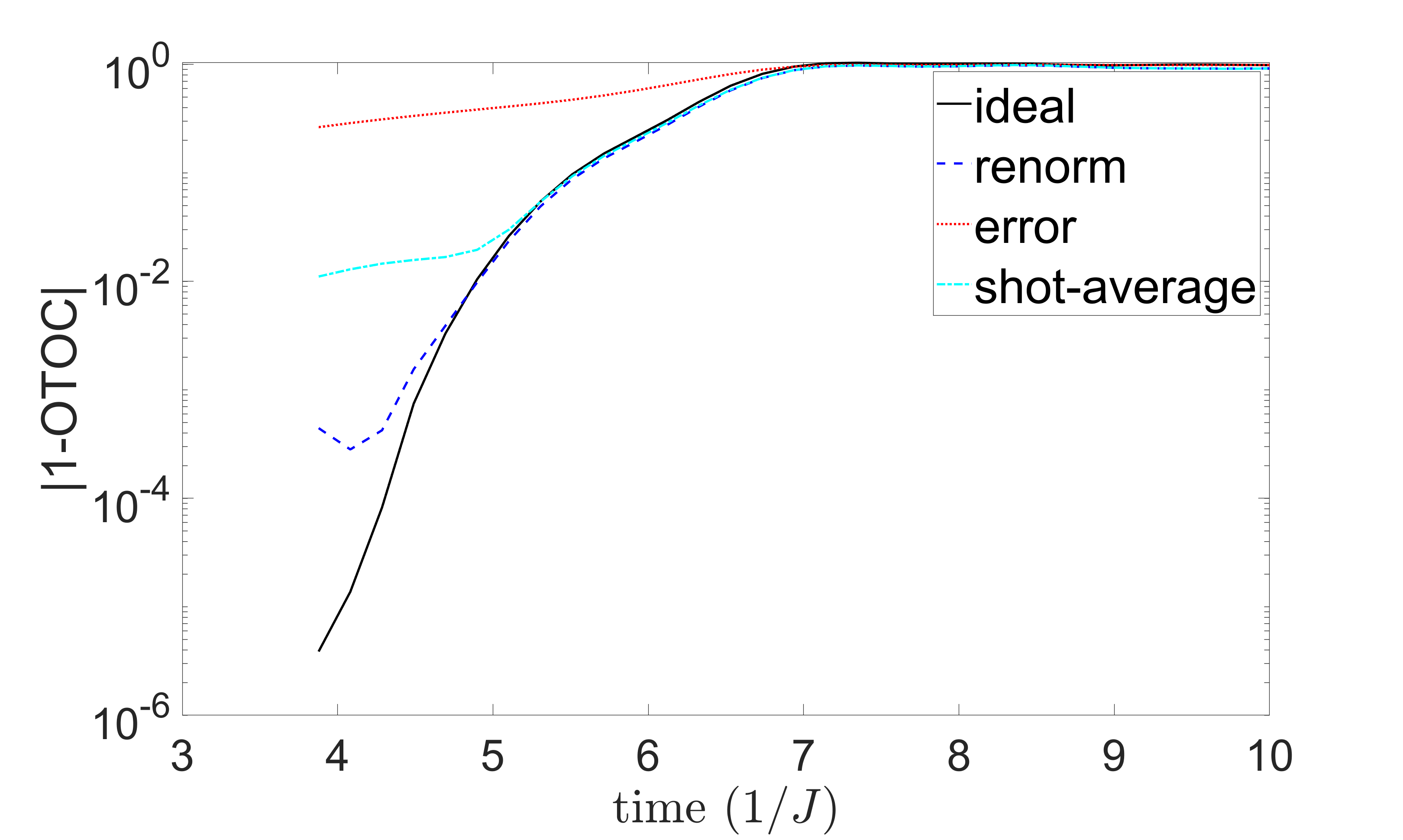}
\caption{\caphead{Shot-to-shot fluctuations:}
Same data as in Figure~\ref{fig:n12eps0p2_floq_manyshot}, on a semilogarithmic plot. }
\label{fig:n12eps0p2_floq_manyshot_log}
\end{figure}

This appendix shows that the renormalization formula also works when the experimental imperfections vary between different experimental shots. This was also the situation considered in the holographic calculation. To minimize computational resources, the numerical results presented are for a Floquet version of the power-law quantum Ising model. Figures~\ref{fig:n12eps0p2_floq_manyshot} and \ref{fig:n12eps0p2_floq_manyshot_log} show
the interferometric renormalization scheme for
a power-law-quantum-Ising Floquet model.
Consider one length-$t$ time evolution.
The Hamiltonian's $\sigma^z$ terms are pulsed on for a short time $dt$;
then the $\sigma^x$ terms are pulsed on for a time $dt$;
then the $\sigma^z$ terms are pulsed on again;
and so on for $t / dt$ time steps. The imperfect time reversal scheme is
the Floquet analog of
the scheme for the Hamiltonian power-law quantum Ising model
(see~\eqref{eq:ImperfectH} and surrounding discussion).
When $\varepsilon=.2$, the ideal and renormalized values are quite close.

On the same figures, we show a shot-to-shot version of the renormalization scheme.\footnote{
Applying the Floquet model
to the shot-to-shot study proves convenient:
Calculating the Floquet model's OTOC requires much less computational time
that calculating a continuous-time model's OTOC.
This computational advantage enables us to average over
many realizations without using too much computer time.}
In practice, an experimenter performs many runs, or shots,
to gather statistics from which to extract the OTOC.
What if the perturbations to the Hamiltonians
vary from shot to shot?
The experimenter can run the experiment many times,
infer a shot-averaged imperfect OTOC,
and infer a shot-averaged imperfect OTOC whose $W= \id$.
The experimenter can divide the former shot-averaged OTOC by the latter.
That this imperfect ratio equals the ideal is unclear.
But the results are surprisingly favorable.

The renormalization formula~\eqref{eq:FInt_Conjecture}
predicts that, for each shot,
\begin{align}
\FInt{t} ( W , V ) \approx \FInt{t} ( \id , V ) F_t.
\end{align}
An experimenter typically cannot measure, in one shot,
all the quantities in this equation.
But $F_t$ is the same for every shot.
Therefore, the shot-averaged quantities (denoted by overlines) obey
\begin{align}
\overline{\FInt{t} ( W , V )} \approx \overline{\FInt{t} ( \id , V )} F_t.
\end{align}
The difficulty has been removed:
The renormalization formula is recast in terms of
shot-averaged quantities, which can be measured experimentally.

Averaging over many shots may be advisable generally.
The number of shots needed depends on
(i) the value of $\varepsilon$ and
(ii) how precisely we want to extract the early behavior.
Figures~\ref{fig:n12eps0p2_floq_manyshot} and \ref{fig:n12eps0p2_floq_manyshot_log}
show averages over just 100 samples.
The ideal and shot-averaged curves
agree reasonably well nonetheless.

\end{appendices}

%
%
\bibliographystyle{h-physrev}
\bibliography{OTOC_bib}


\end{document}